\documentclass[useAMS,usenatbib]{mn2e}

\usepackage{aas_macros}
\usepackage{graphicx} 
\usepackage{amssymb} 
\usepackage{amsmath}

\newcommand{\hMpc}{{\>h^{-1}\rm  Mpc}}

\def\msun{\mbox{$M_\odot$}}

\def\ms{\mbox{$M_{\ast}$}}
\def\mg{\mbox{$M_{g}$}}
\def\mh{\mbox{$M_{h}$}}
\def\mtran{\mbox{$M_{\rm tran}$}}
\def\fg{\mbox{$f_{g}$}}
\def\muhalo{\mbox{$\mu_{\rm halo}$}}
\def\mustar{\mbox{$\mu_*$}}

\def\lcdm{\mbox{$\Lambda$CDM}}

\def\fmerg{\mbox{$f_{\rm merg}$}}

\title[The growth of galactic bulges through mergers in $\Lambda$CDM haloes revisited]
{The growth of galactic bulges through mergers in $\Lambda$CDM haloes revisited. I. Present-day properties}
\author[Zavala et al.] {\parbox{17.5cm}{ Jesus
          Zavala$^{1,2,3}$\thanks{CITA National Fellow, e-mail: jzavalaf@uwaterloo.ca}, Vladimir Avila-Reese$^{4}$, 
          Claudio Firmani$^{4,5}$ and Michael Boylan-Kolchin$^{6}$ 
        }\vspace{0.3cm}\\ 
        $^1$Department of Physics and Astronomy, University of Waterloo, Waterloo, Ontario, N2L 3G1, Canada\thanks{Current affiliations}\\
        $^2$Max-Planck-Institut f\"{u}r Astrophysik, Karl-Schwarzschild-Stra\ss{}e 1, 85740 Garching bei
        M\"{u}nchen, Germany\\
        $^3$Perimeter Institute for Theoretical Physics, 31 Caroline St. N., Waterloo, ON, N2L 2Y5, Canada$\dagger$\\
        $^4$Instituto de Astronom{\' i}a, Universidad Nacional Aut{\' o}noma
        de M{\'e}xico, A.P. 70-264, 04510, M{\'e}xico, D.F., M{\'e}xico\\
      $^5$INAF-Osservatorio Astronomico di Brera, via E.Bianchi 46, I-23807 Merate, Italy\\
    $^6$Center for Cosmology, Department of Physics and Astronomy, 4129 Reines Hall, University of California, Irvine, CA 92697, USA}

\begin{document}



\maketitle

\label{firstpage}

\begin{abstract}
We use the combined data-sets of the Millennium I and II
cosmological simulations to revisit the impact of mergers in the growth of 
bulges in central galaxies in the \lcdm\ scenario. We seed galaxies 
within the growing CDM haloes 
using semi-empirical relations to assign stellar and gaseous masses, and an analytic 
treatment to estimate the transfer of stellar 
mass to the bulge of the remnant after a galaxy merger. We find that this
model roughly reproduces the observed correlation between the bulge-to-total ($B/T$) mass 
ratio and stellar mass (\ms) in present-day central galaxies as well as their observed 
demographics, although low-mass $B/T<0.1$ (bulgeless) galaxies might be scarce
relative to the observed abundance.
In our merger-driven scenario, bulges have a composite stellar population made of
(i) stars acquired from infalling satellites, (ii) stars transferred from the primary disc due to
merger-induced perturbations, and (iii) newly formed stars in starbursts triggered by mergers. 
We find that the first two are the main channels of mass assembly, 
with the first one being dominant for massive galaxies, creating large bulges with different stellar 
populations than those of the inner discs, while the second is dominant for intermediate/low-mass galaxies and creates small bulges with 
similar stellar populations to the inner discs.
We associate the dominion of the first (second) channel to classical (pseudo) bulges, and compare
the predicted fractions 
to observations. We emphasize that our treatment does not include other mechanisms of bulge growth such as
intrinsic secular processes in the disc or misaligned gas accretion.
Interestingly, we find that the evolution of the
stellar and gaseous contents of the satellite as it spirals towards the central galaxy is a key ingredient in 
setting the morphology of the remnant galaxy, and that a good match to the observed bulge demographics
occurs when this evolution proceeds closely to that of the central galaxy. 

\end{abstract}


\section{Introduction}

In the $\Lambda$ Cold Dark Matter (\lcdm) cosmogony, galactic discs 
generically form inside the growing CDM haloes, while spheroids (bulges and elliptical galaxies) 
are thought to grow through mergers, by disc secular internal 
processes, and/or by mis-aligned/perturbed infalling gas \citep[][and references therein]{Mo+10}. 
Several arguments collected over the past decades suggest that major 
mergers of galactic discs produce prominent spheroids that are gravitationally supported by 
random motions. Spheroids built this way might constitute the present population of elliptical and classical 
bulge-dominated galaxies \citep[e.g.,][]{Toomre1977,White78,Gerhard81,
Schweizer82,Negroponte+83,Barnes88,Hernquist92}.
The demographics of these (early-type) galaxies are thus expected to be 
tightly connected to their past galaxy merger activity, 
which is linked to the major merger history of their haloes. 
An open question is whether the observed morphological distribution of galaxies, globally measured by the 
bulge-to-total mass ($B/T$) ratio, and its evolution with redshift, is consistent
with the halo merger rates predicted by the \lcdm\ cosmogony.

In order to answer this question, it is necessary to understand how tight is the relation 
between the growth of spheroids and the merger histories of their host haloes.
The merger-driven growth of spheroids along the evolutionary path of galaxies 
is more complex than assuming that this growth is proportional to the history of halo mergers
with a mass ratio above a given threshold.  
This caused by both, the non-linear relation between halo
mass and stellar mass\citep[e.g.,][]{Conroy+09,Moster-10,Behroozi-Conroy-Wechsler-10, FA10,Leauthaud+2012,Yang+2011} 
and the connection between progenitor gas fractions and spheroid 
formation in mergers \citep[e.g.][]{Barnes-Hernquist-96,Springel-Hernquist-05,Governato-09,
Hammer+2009,Hopkins-09b,Stewart-09}.

During a merger, the bulge in the primary galaxy may grow not only by acquiring 
stars from the merging satellite (the secondary), but also by violently relaxed stars transferred 
from the primary disc, and by newly created stars formed in starbursts. 
In \citet[][hereafter H09a]{Hopkins-09a}, the authors developed analytic prescriptions to describe 
these physical processes with the aim of predicting
the amount of stellar material that is finally deposited into
the bulge of the remnant. The parameters of these prescriptions were calibrated to roughly 
agree with the outcome of a large suite of full hydrodynamical simulations. 

Numerical simulations \citep[e.g.][and references therein]{Athanassoula2005,Avila-Reese+2005,Combes2009}
show that bulges can also grow through intrinsic secular transport of angular momentum 
and dynamical heating of the stellar disc, giving rise to the so-called pseudobulges. In contrast with 
classical bulges, pseudobulges have more rotational support and share some of the properties of the inner discs
\citep[for reviews see][]{Kormendy-Kennicutt-04,Fisher+2008}. Demographics of bulge types
in the very local environment show that low-mass galaxies have small $B/T$ ratios ($<0.2$) with
spheroids that mostly belong to the pseudobulges category. This has been raised as a potential issue for 
\lcdm~which would seemingly predict higher fractions of merger-driven bulges, 
that typically have higher $B/T$ ratios and are of the
classical type (\citealt{Kormendy+2010,Fisher+2011}, hereafter FD11, see also \citealt{Weinzirl_2009}, hereafter W+09).
However, when the stellar mergers have a low mass ratio $<0.1$ (see Section 3.2), 
the bulge may actually be predominantly populated by stars in the primary disc that are transferred to the centre and 
dynamically heated by instabilities induced by the merger. This is the dominant
effect in this kind of merger rather than the addition of stars coming from the secondary.
It is thus possible that bulges formed in minor/minuscule stellar mergers
look like pseudobulges. 

Several studies have aimed to establish the connection between the halo merger 
history with the final galaxy $B/T$ ratio (morphology). Some of these works are
based on the semi-empirical halo occupation framework  \citep{Stewart-09,Hopkins-09b,Hopkins-10a},
while others are based on semi-analytic models (SAMs) \citep{Khochfar-06,Parry+2009,Benson+2010,deLucia+2011,Fontanot+2011}. 
The most general results of these works are that: (i) the mapping of halo-halo 
mergers to stellar galaxy-galaxy mergers is far from linear and strongly depends on mass and 
redshift, (ii) the inclusion of the galaxy gas content in mergers significantly reduces the 
final $B/T$ ratio, specially for low-mass galaxies and at higher redshifts, 
and (iii) the $B/T$ ratio predicted  in the \lcdm~ scenario increases with stellar mass in a similar
way as observations, although there seems to be fewer predicted bulgeless galaxies than observed. 

Our goal is to revisit the merger-driven bulge formation in the context of the \lcdm\ cosmogony
and to compare the results with the observed bulge demographics. Here, we focus 
only on present-day {\it central} galaxies, although the population of satellites is considered along
the evolution of central galaxies. The backbone of our model is an approach where, along
the mass aggregation histories (MAHs) of haloes taken from the Millennium Simulations 
\citep[MS,][]{Springel-05,Boylan-Kolchin-09}, the galaxy stellar and gas masses are assigned according to 
to semi-empirically inferred \emph{average} $\ms(\mh,z)$ and $\mg(\ms,z)$ relations. 
The galaxy evolutionary tracks are quite different to the halo 
MAHs and are such that 
at each $z$ there is a transition stellar mass, $\mtran(z=0)\approx 
2\times 10^{10}$ \msun, above which the \ms\ growth has halted and below which \ms\ 
is actively growing \citep{FA10}. The former happens at 
earlier epochs for more massive galaxies (archaeological downsizing) and for the latter,
less massive galaxies are more active at late epochs (downsizing in specific 
star formation rate).
Note that our approach is different to 
that of SAMs since instead of modeling the complex galactic physics, the main 
galaxy properties are assigned at each epoch according to empirical relations.  

The paper is organised as follows. In Section \ref{sect_2}  
we describe the way we estimate the {\it central} halo-halo merging time, and
our semi-empirical approach to seed baryonic central galaxies into the growing progenitors of the 
halo population. 
In \S\S \ref{galaxies}, we present the stellar merger fractions as a 
function of mass and redshift and compare them with recent observational estimates.
In \S\S \ref{growth_bulges_section}, we give our main predictions related to: (i) the growth of bulges
as a function of mass, and the contribution to this growth from the different mechanisms of bulge 
assembly; and (ii) the $B/T$ demographics of central galaxies at $z=0$.
In \S\S \ref{comp_obs}, our predictions are compared with current observational results.
Finally, a summary and our conclusions are given in Section \ref{sec_concl}.

\section{Semi-empirical model of bulge growth}\label{sect_2}

\subsection{N-body simulations}\label{sims}

We use the combined data-sets of
the Millennium (MS-I) and Millennium II (MS-II) simulations
\citep{Springel-05,Boylan-Kolchin-09} that 
share the same particle number ($2160^3$)
and were done on the context of a WMAP1 cosmology with parameters: 
$\Omega_m=0.25$, $\Omega_{\Lambda}=0.75$, $h=0.73$,
$\sigma_8=0.9$ and $n_{s}=1$; where $\Omega_m$ and $\Omega_{\Lambda}$ are the
contribution from matter and cosmological constant to the mass/energy density
of the Universe, respectively, $h$ is the dimensionless Hubble constant
parameter at redshift zero, $n_s$ is the spectral index of the primordial
power spectrum, and $\sigma_8$ is the rms amplitude of linear mass fluctuations
in $8\hMpc$ spheres at redshift zero. The MS-II has a box size ($L=137$~Mpc
on a side) that is 5 times smaller than the one of the MS-I, and thus it has a mass
resolution limit 125 times smaller: $9.4\times10^6\msun$. Combining both
simulations we can follow up to high redshift ($z\sim10$) the merger and
accretion histories of haloes having a wide mass range at $z=0$: $10^{10}-10^{15}\msun$. 

We should note that the WMAP1 cosmological parameters used in the simulations are different
to those currently preferred by the 7-year WMAP results. In particular, $\sigma_8$ is lower and $\Omega_m$ 
is higher in WMAP7 which produces compensating effects in the abundance and clustering of dark matter haloes.
\citet{Guo+12} studied the impact of both cosmologies on galaxy formation using
a SAM and found that both predict a similar evolution
in the galaxy properties since $z=3$ with only a slightly lower autocorrelation function at separations 
$\lesssim 1$ Mpc, particularly for lower masses, and also a lower fraction of satellites. Hence, a slightly lower
merger rate between satellites and centrals (producing remnants with lower $B/T$ ratios) would be expected if a WMAP7 
cosmology was used instead of the one we adopt here. The difference however is not expected to be significant.

\subsection{Subhalo merger histories and central merger times}\label{times_section}

Since our goal is to analyse the impact of mergers in the growth of the spheroidal
component (bulges) of  {\it central galaxies}, we extract the
subhalo merger histories of the principal branches of a population of
{\it main subhaloes} defined at $z=0$. A main subhalo is the most massive structure
within the hierarchy of subhaloes of a friend-of-friends (FOF) halo, and henceforth we refer to it as a distinct or main halo.
We explicitly reject all subhaloes that coexist within a main halo at $z=0$ as being part of its merger 
history; although they will likely merge with the central object in the future, these subhaloes have no impact for
the central bulge at $z=0$. We also discard
mergers within progenitors not associated with the
main branch of the host halo. By following this procedure, 
our analysis is closer to the actual
merger history of a population of central galaxies at $z=0$ than an
alternative method based on FOF halo merger histories. 

We have randomly selected two samples of main haloes at $z=0$, 
having 1347 and 1500
members with masses larger than $1.2\times10^{12}\msun~(10^3~{\rm
particles})$ and $9.4\times10^{10}\msun~(10^4~{\rm particles})$
for the MS-I and MS-II, respectively. Both samples (properly
normalized to account for the fractional volume they cover relative to the
whole simulation boxes) follow the mass function of the full halo population. 
For each halo in the samples, we obtain their subhalo merger history
using the MS on-line databases\footnote{http://gavo.mpa-garching.mpg.de/Millennium/}. 
For details in the construction of these merger trees see \citet{Springel-05}.

A given merger event is defined by three epochs:
\begin{enumerate}

\item the start of the merger ($z_{\rm start}$), i.e., when the subhalo was part of an 
independent FOF halo for the last time. The
halo mass ratio of the merger is defined by using the subhalo and main halo masses
at this time, $\muhalo=M_{\rm sub}(z_{\rm start})/M_{h}(z_{\rm start})$; 

\item the ``dissolution'' of the subhalo ($t_{\rm diss}$), i.e., when the merged subhalo at time $t_i$ can no longer be 
resolved as an independent self-bound structure at the following time $t_{i+1}$; 

\item the coalescence of the subhalo with the centre of the main halo ($t_{\rm end}$).  
To compute this time, we adopt a dynamical friction time formula just after the subhalo 
has been dissolved in episode (ii) \citep{Binney_1987}:
\end{enumerate}

\begin{equation}\label{merge_time}
t_{\rm df}=\alpha_{\rm fric}(\Theta_{\rm orb})\frac{V_{\rm vir}r_{\rm sub}^2}{G m_{\rm sub}{\ln }~\Lambda},
\end{equation}
where $\alpha_{\rm fric}(\Theta_{\rm orb})$ encloses information on the subhalo orbit, $V_{\rm vir}$ is
the virial velocity of the host, $m_{\rm sub}$ and $r_{\rm sub}$ are the mass and position of the subhalo relative 
to the host just before dissolution, and ${\rm ln}~\Lambda=(1+M_{\rm vir}/m_{\rm sub})$ is the Coulomb
logarithm with $M_{\rm vir}$ the virial mass of the host. We take $\alpha_{\rm fric}(\Theta_{\rm orb})=1.17 \eta^{0.78}$
\citep{Boylan-Kolchin-Ma-Quataert-08}, where $\eta=j/j_c(E)$ is the orbital
circularity of the subhalo relative to the halo centre\footnote{The subhalo 
has specific angular momentum $j$ and energy $E$, and $j_c(E)$ is the
specific angular momentum of a circular orbit with the same energy and with a
radius $r_c(E)$.}. 
The final cosmic time of the halo-halo central merger, $t_{\rm end}$, 
is the sum of $t_{\rm diss} + t_{\rm df}$;  we consider that $t_{\rm end}$ is a good approximation to the actual 
galaxy-galaxy merger epoch.

\subsection{Galaxy occupation}\label{semi_section}

Although the merger histories of haloes may provide a basis for determining the 
morphology of their central galaxies, the impact of a merger 
does not depend directly on the halo mass ratio at the start of the
merger but rather on the central dynamical masses (inner dark matter, gas and stars) 
that interact in the final stages of the galactic merger. 

To follow the stellar and gas mass assembly
of galaxies inside the main haloes, as well as the
processes that affect the gas and stars during mergers,
we use a semi-empirical approach close to that one in \citet[][]{Hopkins-09b,Hopkins-10a}. 
This approach yields stellar mass assembly histories that are consistent, 
by construction, with observational trends. 
In Appendix \ref{semi_app} we describe in detail this method and its implementation. In summary, for each 
present-day main halo in our MS samples we do the following:

\begin{enumerate}

\item extract the main branch of its merger tree; 

\item seed a central galaxy, the primary, at $z_{\rm seed}=3$ with
stellar and gas masses given by the semi-empirical relations 
$M_\ast(M_h,z)$ and $M_g(M_\ast,z)$ (Appendix \ref{semi_app_1});

\item identify $z_{\rm start}$ for
each halo merger along the main branch ($z_{\rm start}\leq z_{\rm seed}$) and
assign a galaxy to the infalling halo, the secondary, according to
$M_\ast(M_h,z_{\rm start})$ and $M_g(M_\ast,z_{\rm start})$;

\item either assume that the stellar and gas content of the secondary does not evolve until final coalescence,
or follow its evolution (the star formation, SF, and SF-driven
outflow processes) by means of semi-analytic recipes (Appendix \ref{semi_app_3});

\item compute the galaxy-galaxy (central halo) merging time, $t_{\rm end} = t_{\rm dis}
+ t_{\rm df}$, using Eq.~(\ref{merge_time}) for $t_{\rm df}$;

\item estimate the bulge and disc masses of the primary galaxy after coalescence at $t_{\rm end}$, 
using physical recipes calibrated by numerical simulations (Appendix \ref{semi_app_2});

\item repeat the process for the central galaxy until reaching $z=0$,
taking care of each merger and updating at each $z$ the properties of the central
galaxy according to the $M_\ast(M_h,z)$ and $M_g(M_\ast,z)$ relations.

\end{enumerate}

Note that in our scheme galaxies are initially pure discs, 
and that the only channel of bulge growth is through mergers.
We begin testing our model by
comparing the evolution of the galaxy merger rate with observational constraints.

\section{Results}

\subsection{Stellar merger fractions}\label{galaxies}

\begin{figure}
\centering
\includegraphics[height=6.5cm,width=8.5cm,trim=1.2cm 0.5cm 0.5cm 0.5cm, clip=true]{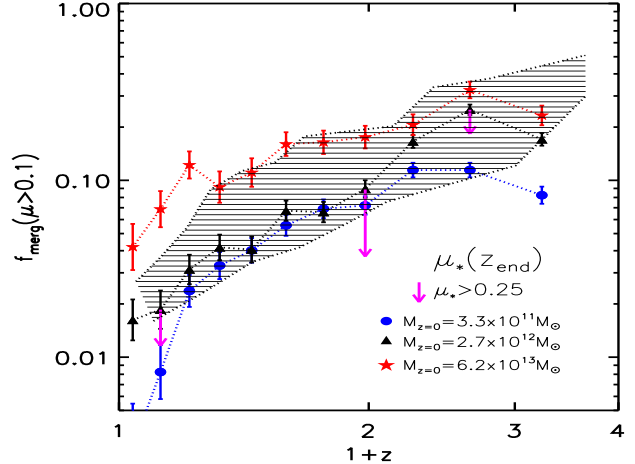}
\caption{Relative fraction of main progenitors having at least a stellar major
merger event with $\mu_\ast>0.1$ as a function of redshift. The time interval
between redshift bins is $1~$Gyr. The fraction is relative to the total
number of main progenitors in a given redshift bin. The
colors (symbols) are for three different halo mass bins at $z=0$ according to the
legend. The low-mass bin is for the MS-II sample, the other two are for the
MS-I sample. The stellar major merger ratio and redshift are defined at the time of
final galaxy coalescence. The secondary galaxies grow according to 
the SAM described in Appendix \ref{semi_app_3}. Stellar and gas masses for each halo are
given by empirical relations described in Appendix \ref{semi_app_1}. 
The shaded region is a compilation of observations by \citet{Hopkins-10a,Hopkins-10b} 
for $\mustar>0.1$. The magenta 
arrows mark the change on $f_{\rm merg}$ for the intermediate mass bin if the major merger 
threshold is increased to $0.25$.}
\label{frac_merger}
\end{figure}

Fig.~\ref{frac_merger} shows \fmerg, the fraction of main halo progenitors that suffer at least 
one central stellar merger with $\mu_\ast>0.1$, relative to the total number 
of main progenitors in a temporal bin of $1$~Gyr, i.e., \fmerg\ is equivalent to the merger rate per halo and per Gyr.
The blue, black, and red 
lines (circle, triangle and star symbols, respectively) are the median values for three different halo mass
bins, defined at $z=0$, centered in $3.3\times10^{11}\msun$, $2.7\times10^{12}\msun$ and
$6.3\times10^{13}\msun$, respectively; the error bars are the corresponding $1\sigma$ scatters of the distributions.  
The mergers (including their stellar mass ratio) are defined at the time when the secondary galaxy finally coalesces with
the central one ($z_{\rm end}$).
In Fig.~\ref{frac_merger} we have actually used $z_{\rm seed}=4$, which is the maximum redshift used to infer the 
halo-to-stellar mass relation.

The figure shows that the fraction of progenitors with on-going mergers grows with $z$, 
which is a trend inherited from the well known growth with $z$ of
the halo-halo merger rates 
\citep[e.g., see right panel, Fig. 3 of][]{Fakhouri-Ma-Boylan-Kolchin-10}. 
We can also see that $\fmerg$ depends on \mh: galaxies in the most massive
haloes have in average a higher fraction of mergers than those in the less
massive ones; at the level of halo mergers, this dependence is quite weak.

\begin{figure}
\centering
\includegraphics[height=6.5cm,width=8.5cm,trim=1.8cm 0.5cm 0.5cm 0.5cm, clip=true]{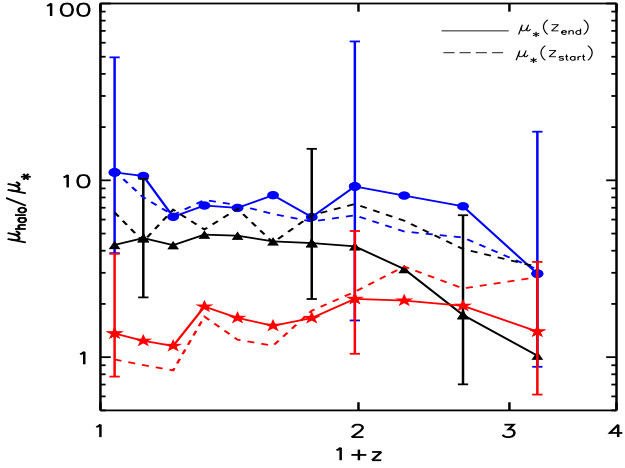}
\caption{Median values and 1$\sigma$ scatter of the distribution of halo-to-stellar merger mass ratios,
\muhalo($z_{\rm start}$)/\mustar($z_{\rm end}$), for the mergers shown in Fig.~\ref{frac_merger}, with 
the same three halo mass bins 
(solid curves from top to bottom are for lower to higher 
masses, respectively) and for the case of evolution in the secondaries. The dashed
lines are the same but with \mustar\ defined at the
beginning of the halo merger.}
\label{mu-ratios}
\end{figure}

It is interesting to remark, as noted elsewhere \citep[][]{Stewart-09,Hopkins-10a}, 
that there is a dramatic change in the merger fraction as a function of mass and $z$ depending
in which mass ratio is used to define the merger. If we use \mustar~instead of \muhalo, there is an overall drop in the fraction 
of main progenitors with major mergers, and the dependence of
$\fmerg$ on mass increases, specially for the largest masses. This is mainly produced by the shape of 
the $M_{\ast}(M_h,z)$ relation: at masses below the knee of this relation, 
a $\mu_\ast(z_{\rm start})= 0.1$ merger is actually related to a $\mu_{\rm halo}(z_{\rm start})=0.36$ merger
(because $M_h\propto M_\ast^{0.44}$), which is more rare than a $1:10$ halo merger; at masses above the knee, 
where $M_h\propto M_\ast^2$, an opposite behavior is in principle expected.
The stellar growth of the merging galaxies from $z_{\rm start}$ until $z_{\rm end}$ also contributes
to the changes in the merger fraction. To give an impression of this,
we show in Fig. \ref{mu-ratios} the median values of 
 \muhalo($z_{\rm start}$)/\mustar($z_{\rm end}$) as a function of $z$ for the mergers corresponding to Fig.~\ref{frac_merger},
in the case where the secondaries evolve according to a simple SAM (see Appendix \ref{semi_app_3}). 
The MS samples were divided into the same three present-day \mh\ bins as in Fig.~\ref{frac_merger}. 
The error bars show the $1\sigma$ scatters. On average, a \muhalo($z_{\rm start}$)=0.1 halo merger 
corresponds to a $\mustar(z_{\rm end})\sim0.01$ stellar merger for low-mass galaxies.  
Interestingly, notice that in our model
of secondary evolution, the \mustar\ ratios are roughly
the same whether they are measured at the beginning ($z_{\rm start}$, dashed lines) or at end of the merger 
($z_{\rm end}$, solid lines). {\it This means that in average, the secondary increases its stellar mass (semi-analytically calculated) 
roughly by the same amount as the primary (semi-empirically assigned) 
during the period from infall to coalescence.}

In the case of the baryonic (stars + cold gas) merger-mass ratio, 
the fraction of haloes having mergers with $\mu_{\rm bar}>0.1$ increases
relative to that based on stellar major mergers. This is because the fraction of cold gas is
significantly high for galaxies with low stellar masses, specially at high redshifts. 
A significant fraction of mergers with \mustar$<0.1$ are actually major baryonic mergers
just because the secondaries have larger cold gas fractions than the
primaries. 
Morphology-based techniques for estimating merger rates are more sensitive
to baryonic than to stellar mergers, and indeed they find higher
merger rates than with pair-based techniques \citep{Lotz+2011}.

The shaded area in Fig.~\ref{frac_merger} encompasses the observed
major-merger fractions compiled 
by \citet{Hopkins-10a,Hopkins-10b}. Most 
of the observational results do not have a well-defined mass selection and are based 
on different merger identification criteria.
Very roughly, the stellar mass of the primary galaxies covers a range $10^{10}-2\times 10^{11}$ \msun\
at low redshift, with a larger minimum mass at high $z$. For measurements based on pair samples (pre-merger), 
$\mu_{\ast}\gtrsim 0.25$, while for measurements based on morphology samples (post-merger), $\mu_{\ast}\gtrsim 0.1$.
The lower (upper) bound 
of the shaded region is dominated by the pair (morphology) samples, and therefore
reflects merger fractions with $\mu_{\ast}\gtrsim 0.25$ ($\mu_{\ast}\gtrsim 0.1$). 

Fig.~\ref{frac_merger} shows that there is a reasonable agreement between predictions
and observations\footnote{We note that 
while our predictions refer to evolutionary tracks of individual galaxies defined at $z=0$,
observations refer to samples above a constant \ms\ at all epochs. For
very massive galaxies, \ms\ changes little since $z\sim2$, but for intermediate and low-mass galaxies, 
\ms\ is smaller at higher redshifts. Hence, the prediction for the merger fraction
above a certain \ms\ threshold, constant at all epochs, would be curves slightly steeper
than the ones shown in Fig.\ref{frac_merger}, particularly for the intermediate and low-mass bins.}. 
If we limit the mergers to \mustar($z_{\rm end}$)$> 0.25$, then the fraction falls
below the lower limit of observational estimates for $z<1$ (magenta arrows  
for the intermediate mass bin where most galaxies have \ms($z=0$)$\sim 1-5 \times 10^{10}$\msun). 
In a recent study based on the 30-band photometric catalogue in COSMOS, complemented with the spectroscopy 
of the zCOSMOS survey to define close pairs, \citet{Lopez-Sanjuan+2012} report the  merger fraction 
evolution ($0.2\lesssim z\lesssim 1$) of massive galaxies, $\ms>10^{11}$\msun (corresponding to $\mh\gtrsim 7\times 10^{12}$\msun\ ), 
for $\mustar>0.1$. The predicted merger fraction evolution is in reasonable agreement with these observations.

\subsection{Merger-induced bulge growth and its dependence on satellite evolution}\label{growth_bulges_section}

After showing that the \lcdm-based galaxy merger fractions (rates) are roughly 
consistent with observations, we concentrate now in the growth of bulges during
mergers through the following processes (see Appendix \ref{semi_app_2}):
\begin{description}
\item[(a)] the total acquisition of stars from the secondary, 
\item[(b)] the violent relaxation and transport to the centre of a fraction of stars in the primary disc, and 
\item[(c)] the newly formed stars in central starbursts produced by a fraction of the gas from the merging galaxies (Eq. A2).
\end{description}

To calculate the amount of stars added to the bulge through processes (a)--(c), we estimate the
dynamical (stars, gas, inner dark halo) masses of the merging galaxies. 
As mentioned above, the empirical $M_{\ast}(M_h,z)$ and $M_g(M_{\ast},z)$ relations
are used for this whenever is possible. These relations hold for the primary at any point before final coalescence 
and shortly afterwards. For the secondary however, they are only
valid before the start of the halo merger. As the secondary spirals inwards towards the centre,
different physical processes alter its stellar and gaseous
contents.  
To study how important some of these processes are, we first analyse the case where the gas and stellar masses
in the secondary are ``frozen'' at the values they had at the time the halo merger started
(\S\S \ref{no_evolution}), and we later move to the more realistic case of stellar and gaseous evolution
in the satellite (\S\S \ref{evolution}).

In order to use both MS data-sets without introducing a bias due to 
numerical resolution, we consider only those mergers with $\mu_{\rm halo}\ge 0.1$ in what follows.
At this level, both simulations capture the majority of the halo mass contributed by
mergers and statistically match each other in their overlapping mass range.
This could raise the concern that by missing all mergers with $\mu_{\rm halo}<0.1$, the growth of
bulges might be affected significantly. However, due to the shape of the $M_{\ast}(M_h,z)$ relation, halo-halo mergers with 
$\mu_{\rm halo} = 0.1$, imply galaxy-galaxy mergers with $\mu_{\ast}\ll0.1$
at low masses and  $\mu_{\ast}\sim 0.1$ at large masses (see  
Fig. \ref{mu-ratios}). Thus, it is only for the most massive systems that some
relatively large stellar mergers are being excluded, which are in any case not very
relevant since the bulk of the bulge is formed during the largest merger 
events (see e.g. Fig. 8 of \citealt{Hopkins-10a}).
For massive haloes, we have nevertheless checked the role of including 
those events with $\mu_{\rm halo}<0.1$, only possible in the MS-II,
and found that the results change very little.
\begin{figure*}
\centering
\includegraphics[height=5.5cm,width=5.5cm,trim=1.2cm 0.5cm 1.0cm 1.0cm, clip=true]{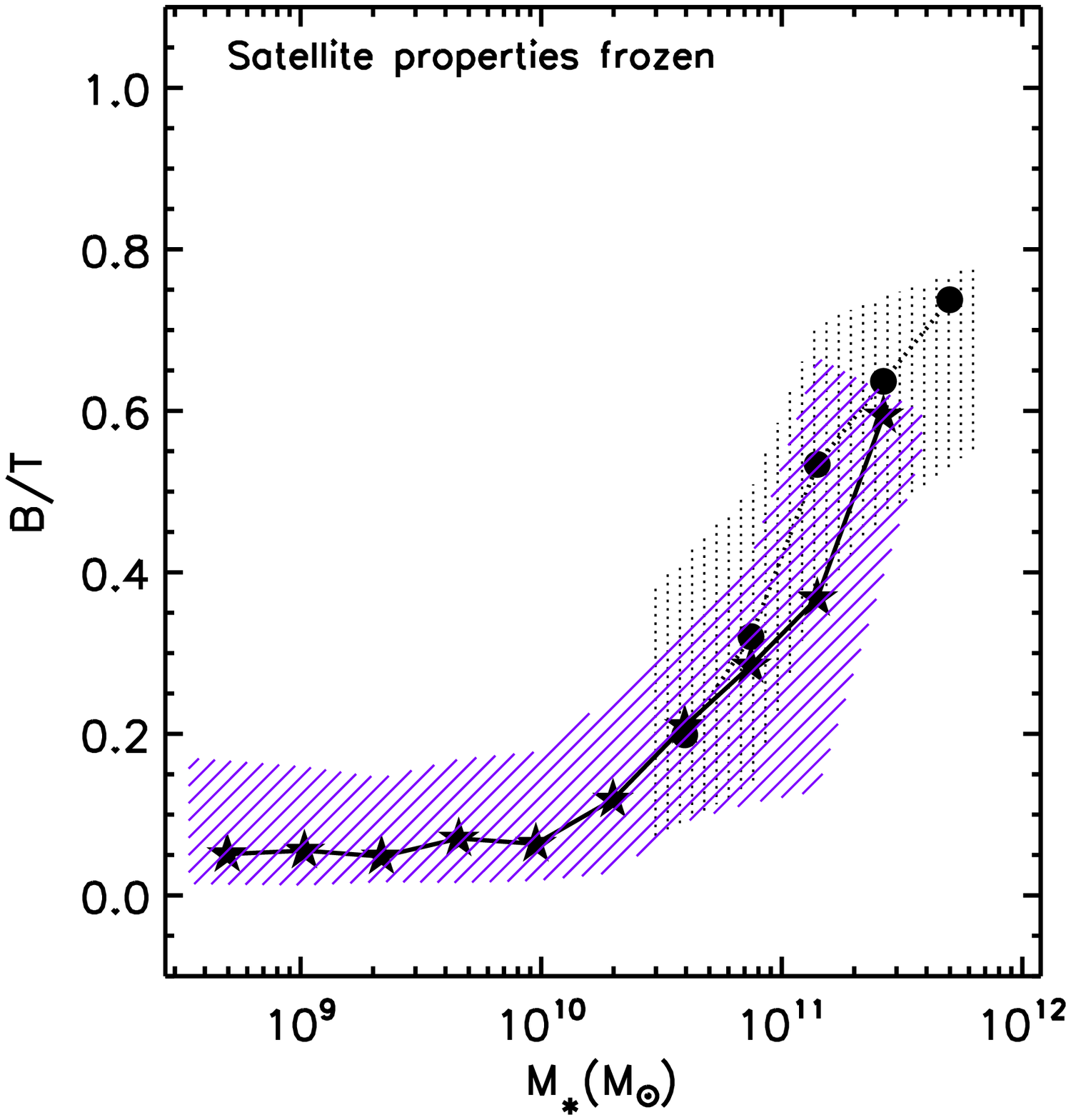}
\includegraphics[height=5.5cm,width=5.5cm,trim=1.2cm 0.5cm 1.0cm 1.0cm, clip=true]{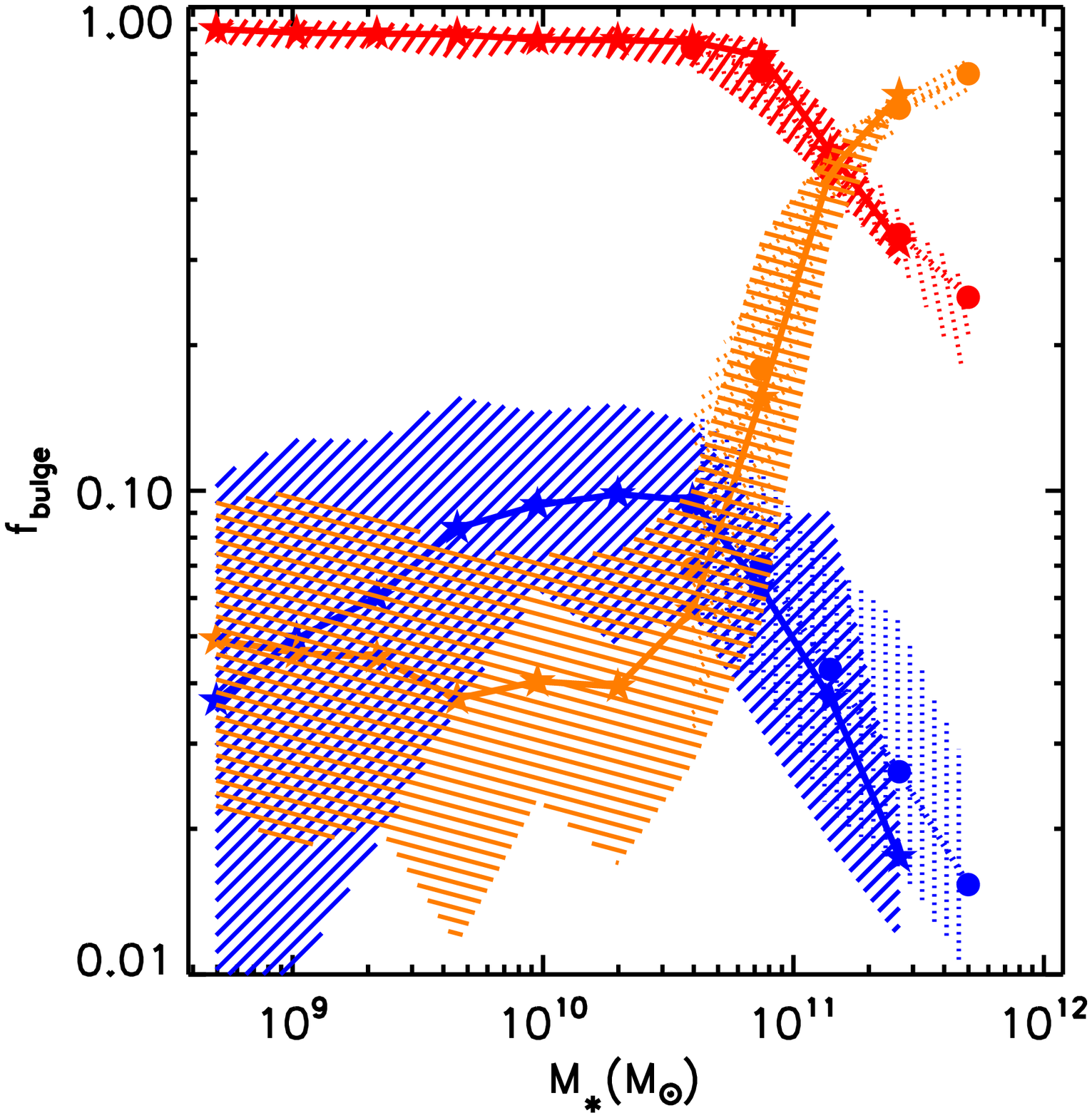}
\includegraphics[height=5.5cm,width=5.5cm,trim=1.2cm 0.5cm 1.0cm 1.0cm, clip=true]{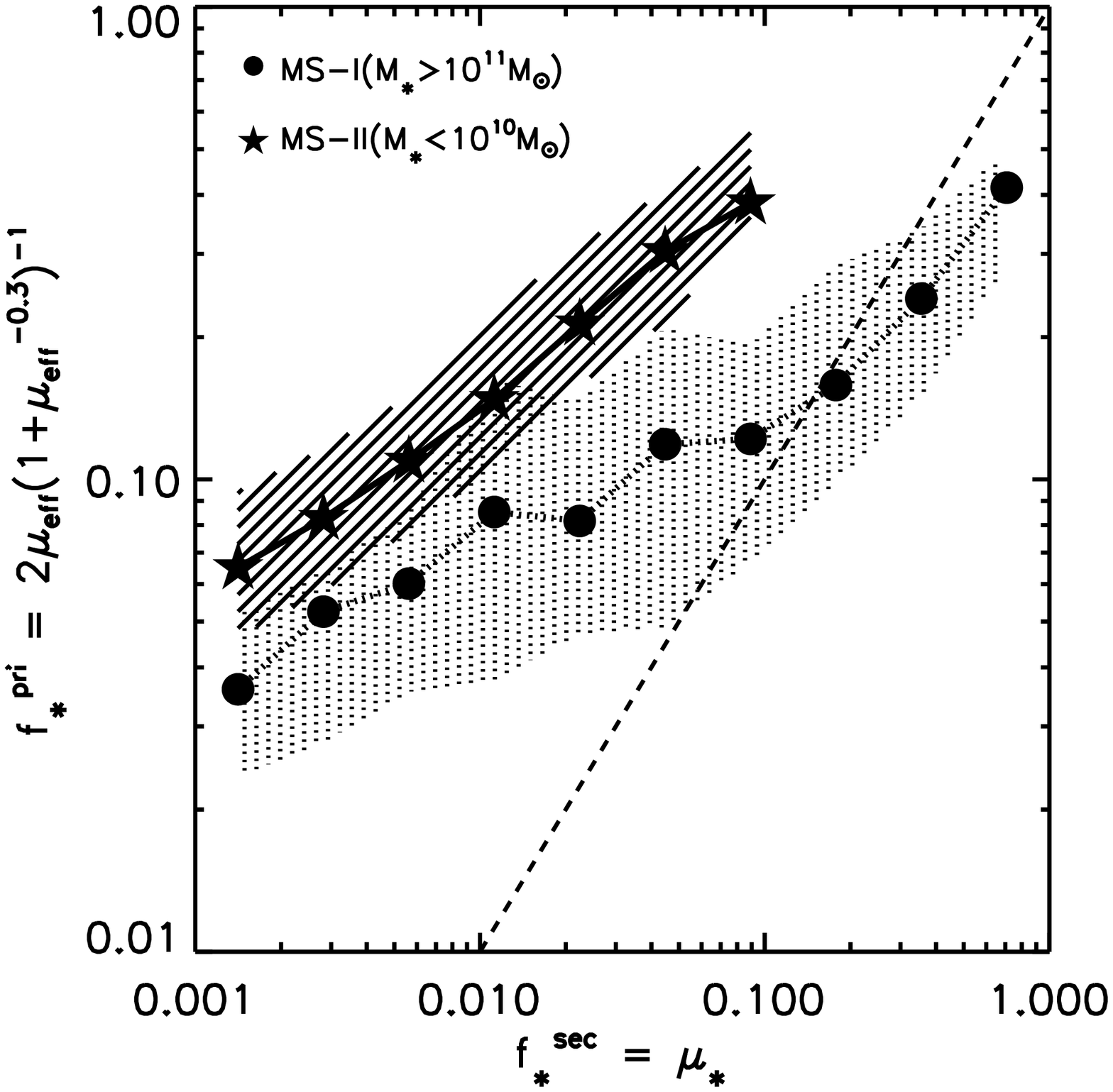}
\includegraphics[height=5.5cm,width=5.5cm,trim=1.2cm 0.5cm 1.0cm 1.0cm, clip=true]{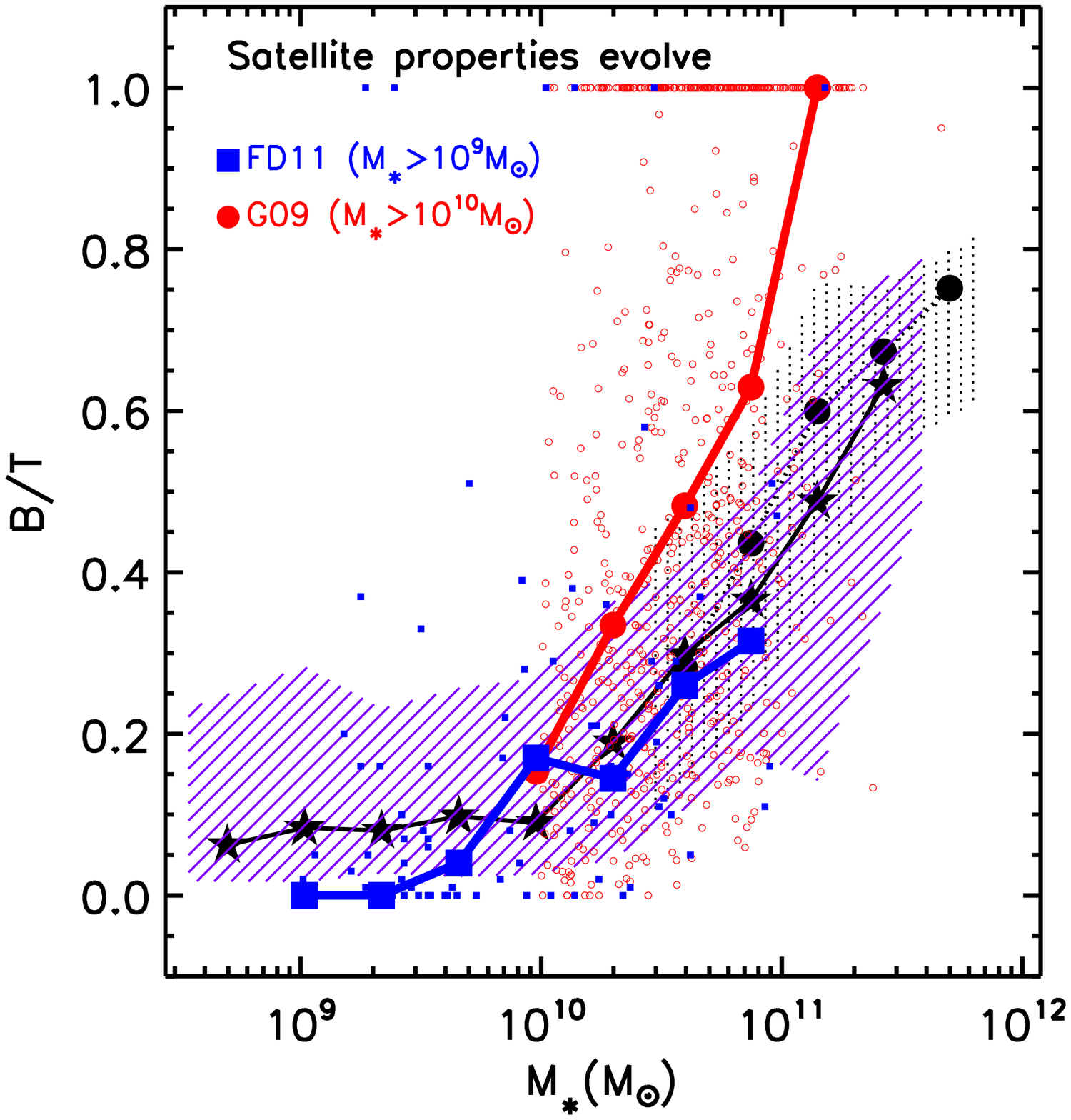}
\includegraphics[height=5.5cm,width=5.5cm,trim=1.2cm 0.5cm 1.0cm 1.0cm, clip=true]{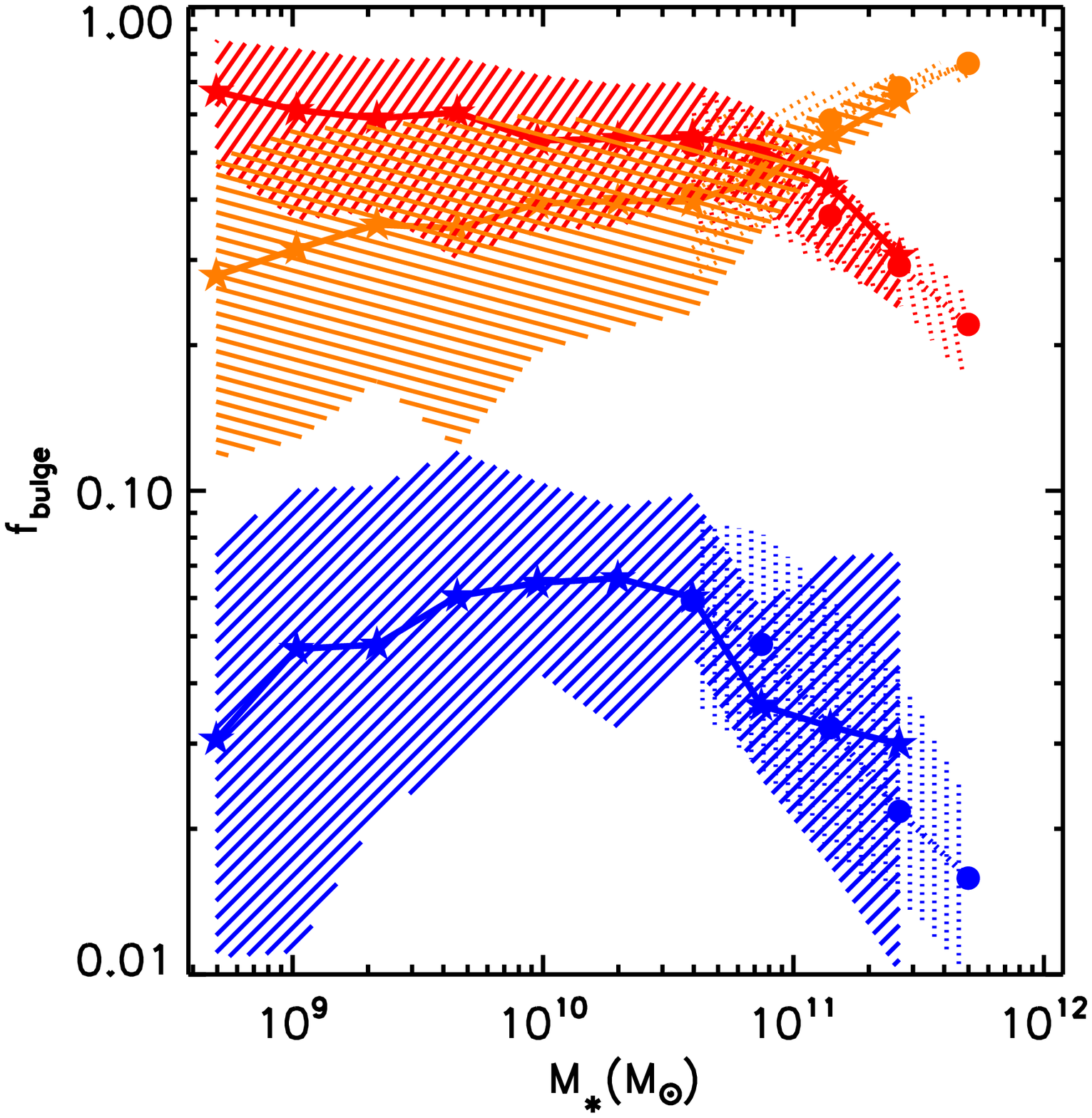}
\includegraphics[height=5.5cm,width=5.5cm,trim=1.2cm 0.5cm 1.0cm 1.0cm, clip=true]{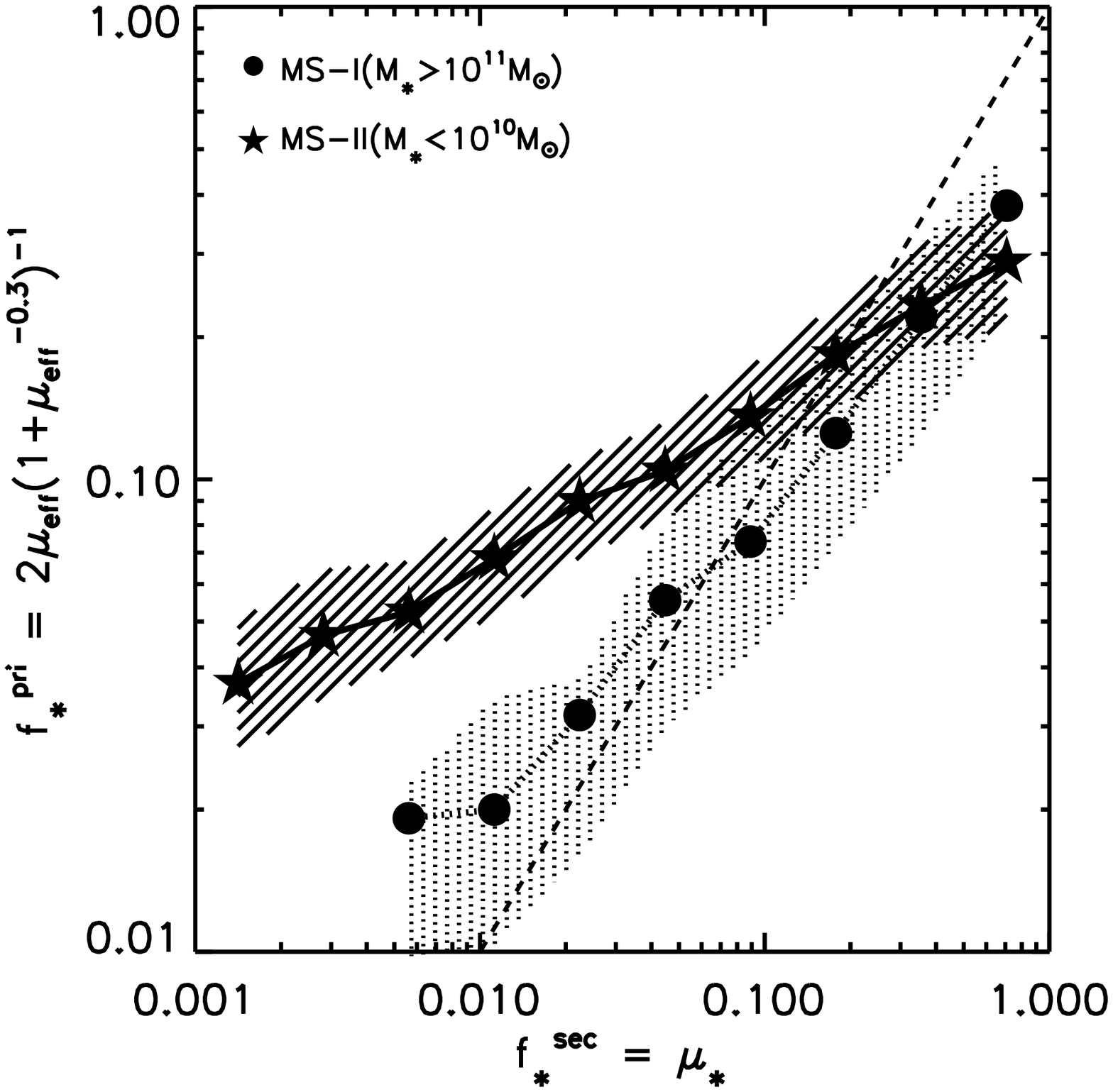}
\caption{{\it Left panels:} Bulge-to-total stellar mass ratio as
a function of $M_{\ast}(z=0)$. {\it Middle panels:} Fraction of the final stellar
bulge mass assembled by the three mechanisms of bulge growth in a merger: stars formed in
starbursts (blue), stars transferred from the primary disc (red), and stars from the secondary 
(orange). {\it Right panels:} Comparison between the fraction of stars in
the primary, $f_\ast^{\rm pri}$, that are transferred to the bulge and the fraction
of stars that are added via the merger of the secondary, $f_\ast^{\rm pri}$; the one to one relation
between these fractions is shown as a dotted line. These fractions are
relative to the mass of the primary just before final coalescence. In all
panels, the symbols and solid lines are the median values 
and the shaded areas contain the $\pm\sigma$
regions of the distribution. Circles (stars) and dotted (full) lines patterns
are for the MS-I (MS-II). In the right panels, the sample of galaxies was
divided in two mass ranges: $M_{\ast}<10^{10}$M$_{\odot}$ (MS-II data only)
and $M_{\ast}>10^{11}$M$_{\odot}$ (MS-I data only). Galaxies were seeded at $z_{\rm seed}=3$ and
only mergers with $\mu_{\rm halo}>0.1$ have been included. The upper row is for the case where 
the fraction of gas and stars in the secondary is the same at the time of final coalescence as it was when
the halo merger started, whereas the lower row is for a model where the gas and stars
in the secondary evolve through SF and SN feedback after
the halo merger starts (Appendix \ref{semi_app_3}). Observational data for two
galaxy samples with $M_{\ast}\geq10^{10}$M$_{\odot}$ from 
G09 and FD11 has been added to the
left panel with red circles and blue squares, respectively. The median of the distribution for the
galaxy samples are shown with solid lines.}
\label{BT_mass}
\end{figure*}
%
\subsubsection{Case 1:  non-evolving secondary} 
\label{no_evolution}

The upper left panel of  Fig.~\ref{BT_mass} shows $B/T$ as 
a function of \ms($z=0$) for the central galaxies seeded
into the MS-I and MS-II haloes (circle and
star symbols, dotted and full-line shaded regions, respectively) in the case where 
secondaries have, at coalescence, the same \ms\ and \mg\ they had when their haloes
became subhaloes (extreme SF quenching). Since mergers
are more frequent in massive galaxies and since the gas fraction is a decreasing
function of \ms, it is expected that massive galaxies will be more prone to 
develop large bulges. 
Several studies in the past have shown the behavior
depicted in Fig.~\ref{BT_mass} for a diverse class of models 
\citep[e.g.][]{Hopkins-09b,Hopkins-10a,deLucia+2011}.
Below a few times $10^{10}$M$_{\odot}$, most galaxies have $B/T\leq0.2$.
This is partially an effect of the high gas fractions
of low-mass galaxies, which make the bulge growth less efficient.

The middle panels of Fig.~\ref{BT_mass} show the contribution  
of the three merger-driven processes of bulge growth mentioned above, 
as a function of \ms. The bulge mass fraction assembled through processes (a), (b), 
and (c) is shown with orange, red and blue, respectively.
In the right panels of Fig.~\ref{BT_mass}, we compare the fraction of stars violently relaxed 
in the primary (which is  a function of the effective dynamical mass ratio,
$\mu_{\rm eff}$, see Eq.~\ref{violent_rel}) to the fraction of stars acquired
from the secondaries (which is equal to the stellar mass ratio $\mustar$) for each merger. We
have divided the events according to the mass of their descendant at 
$z=0$: massive galaxies with $M_{\ast}>10^{11}$M$_{\odot}$ (circles and dotted region) and low-mass galaxies with
$M_{\ast}<10^{10}$M$_{\odot}$ (stars and dashed region). 

In the case of no satellite evolution, the bulges of $\ms\lesssim 10^{11}$ \msun\ galaxies grow overwhelmingly
through stars transferred from the primary rather than by stars acquired from
the secondary with $\mu_\ast<0.1$ 
(i.e., the stellar mergers are essentially always minor).
This is mainly because the secondaries are gas dominated in most of the cases.
For the same reason, we see in the middle panel (upper row) that
the contribution from starbursts (blue) for galaxies of this mass is also more important than the one from stars
in the secondaries (orange).
Nevertheless, starbursts never contribute more than $\sim 10\%$ to the present-day mass bulge, and
they are essentially negligible for massive galaxies due to their very low gas content.

For galaxies with $\ms> 2\times 10^{11}$ \msun, it is possible to have mergers with
\mustar$>0.1$, and when these mergers reach \mustar$\sim0.2$, then the dominant channel of 
bulge growth is stars directly added from the secondary. Thus, the difference between galaxies 
of different mass in the right panel of Fig.~\ref{BT_mass} explains clearly the trends seen in the 
middle panel between the red and orange distributions. {\it This is ultimately connected to the shape of 
the $M_{\ast}(M_h)$ relation.}

\subsubsection{Case 2: evolving secondary}
\label{evolution}

The effect of including SF and SN feedback in the
evolution of the secondary in its travel to the centre of the primary (Appendix \ref{semi_app_3})
is explored in the lower panels of Fig.~\ref{BT_mass}. These processes augment the stellar mass 
of the secondary and reduce its gas fraction. They are of course more relevant when the supply of
gas at the start of the merger is high and thus, they considerably affect
low-mass galaxies and are nearly irrelevant for massive galaxies. In what follows we take 
this model as our fiducial one. Note that this case also implies some level of SF quenching since
once a galaxy becomes satellite, there is no newly accreted gas.

The $B/T$ ratio of low-mass galaxies is on average slightly higher in this case. 
Continuous SF increases the stellar mass of the secondary at  
the time of coalescence relative to the start of the merger,
making the contribution from stars in the secondaries to the bulge growth 
much more significant (orange region, middle panels of Fig.~\ref{BT_mass}). 
Galaxy outflows from SN feedback attenuate slightly this
effect by removing gas from low-mass secondaries. 
On the other hand, the other two mechanisms (b) and (c),   
are diminished because $\mu_{\rm eff}\sim \mu_{\rm bar}$ is reduced due to the loss of gas 
(compare right panels of Fig.~\ref{BT_mass}).

Bulge stars formed in merger-induced starbursts contribute on average less than $\sim 5\%$
to the total masses of present-day bulges, the largest contribution being in galaxies with masses
$\ms\approx 2\times 10^{10}$ \msun.  
H09a and \citet{Hopkins-10a} report a larger contribution, particularly for low-mass
galaxies, reaching $\sim40\%$ (see Fig. 14 of \citealt{Hopkins-10a}).
This difference is partly caused by the model in \citet{Hopkins-10a} being closer to our previous model
where the gas fractions are not allowed to decrease while the satellite merges. Moreover, we also
speculate that an additional cause of the discrepancy is the
different treatment of the extrapolation of the gas fraction \fg$=M_g/M_\ast$ to lower masses 
and higher redshifts: we 
have put an upper limit on \fg~based on the maximum observed value in the compilation of observations given
by \citet{Stewart-09}: $\fg\leq100$, whereas \citet{Hopkins-10a} uses a less restrictive limit 
(see Appendix \ref{semi_app_1}). We found that the latter case
indeed creates a larger contribution from starbursts 
in low-mass galaxies, although not to the level reported in \citet{Hopkins-10a}.  It
 also creates more destructive mergers (more massive
bulges), simply because there is more
SF prior to the final coalescence. Because of this, it is hard to increase the contribution
from starbursts while at the same time keeping the $B/T$ distribution consistent with observations.

In spite that the contribution of stars from the secondaries is increased 
compared to the case of no satellite evolution, the contribution from stars violently relaxed in the primary
continues to be dominant for $\ms\lessapprox 10^{11}$ \msun. 
For $\ms\approx 10^9, 3\times 10^{10}$, and  $10^{11}$ \msun, on average 60\%, 55\%, and 45\% 
of the final bulge mass was accreted through this channel, respectively.  {\it This implies that the smaller the galaxy, 
the more their bulges share properties of their stellar population with their discs}.
Notice that this is an important result since it suggests that bulge growth due to secular evolution and 
that induced by mergers cannot be distinguished only by looking at the similarity of the stellar populations of the
disc and the bulge, as is sometimes assumed.
There are however some small galaxies with bulges mostly assembled
by stars from the secondaries; we can see that this happens when the mergers with $\mu_{\ast}\gtrsim 0.25$ dominate
(right panel of Fig. \ref{BT_mass}, lower row).

\subsubsection{Comparison with observations: $B/T$ demographics}\label{comp_obs}

\begin{figure*}
\centering
\includegraphics[height=6.5cm,width=7.5cm,trim=1.4cm 0.5cm 1.1cm 1.0cm, clip=true]{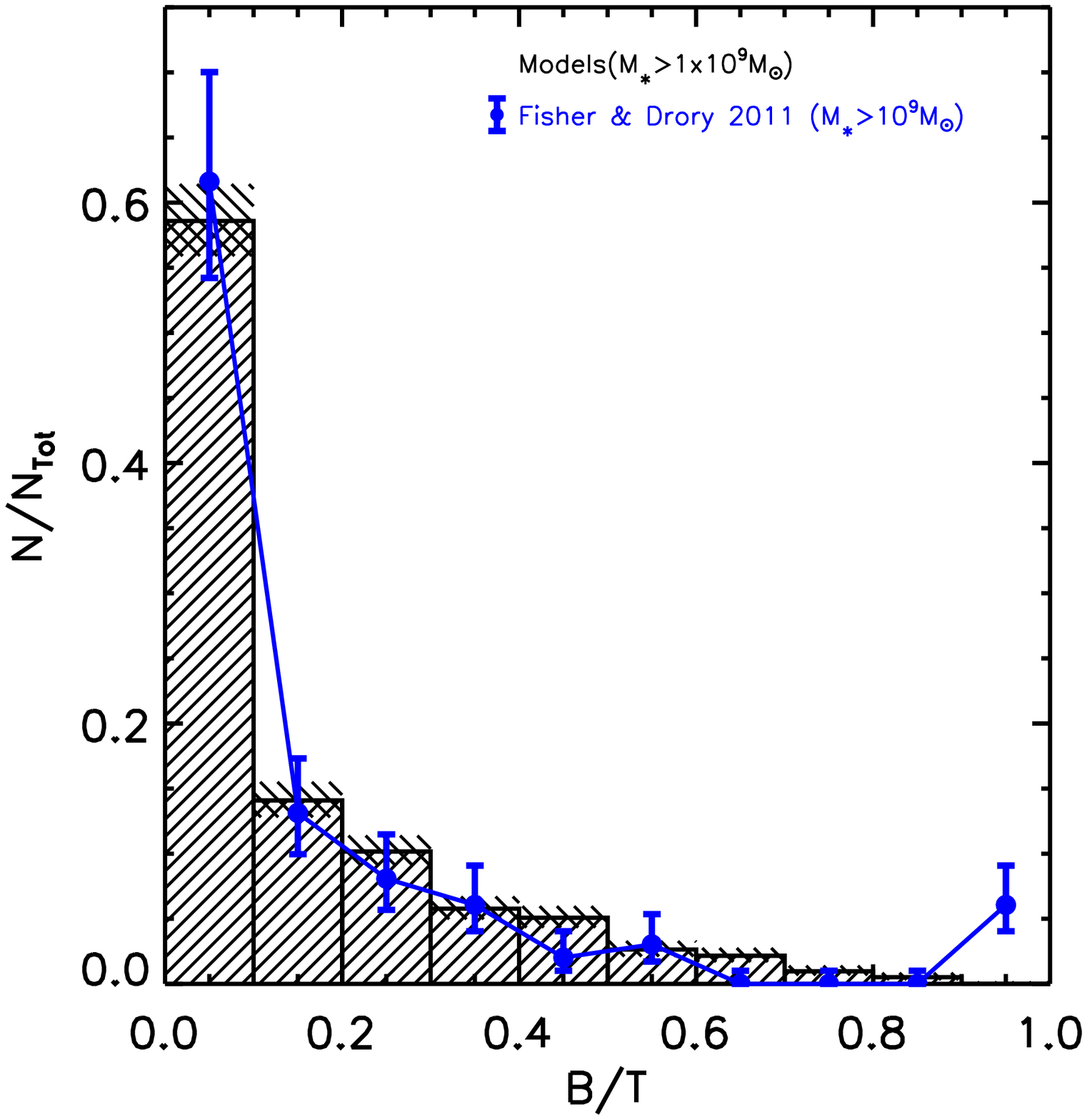}
\includegraphics[height=6.5cm,width=7.5cm,trim=1.4cm 0.5cm 1.1cm 1.0cm, clip=true]{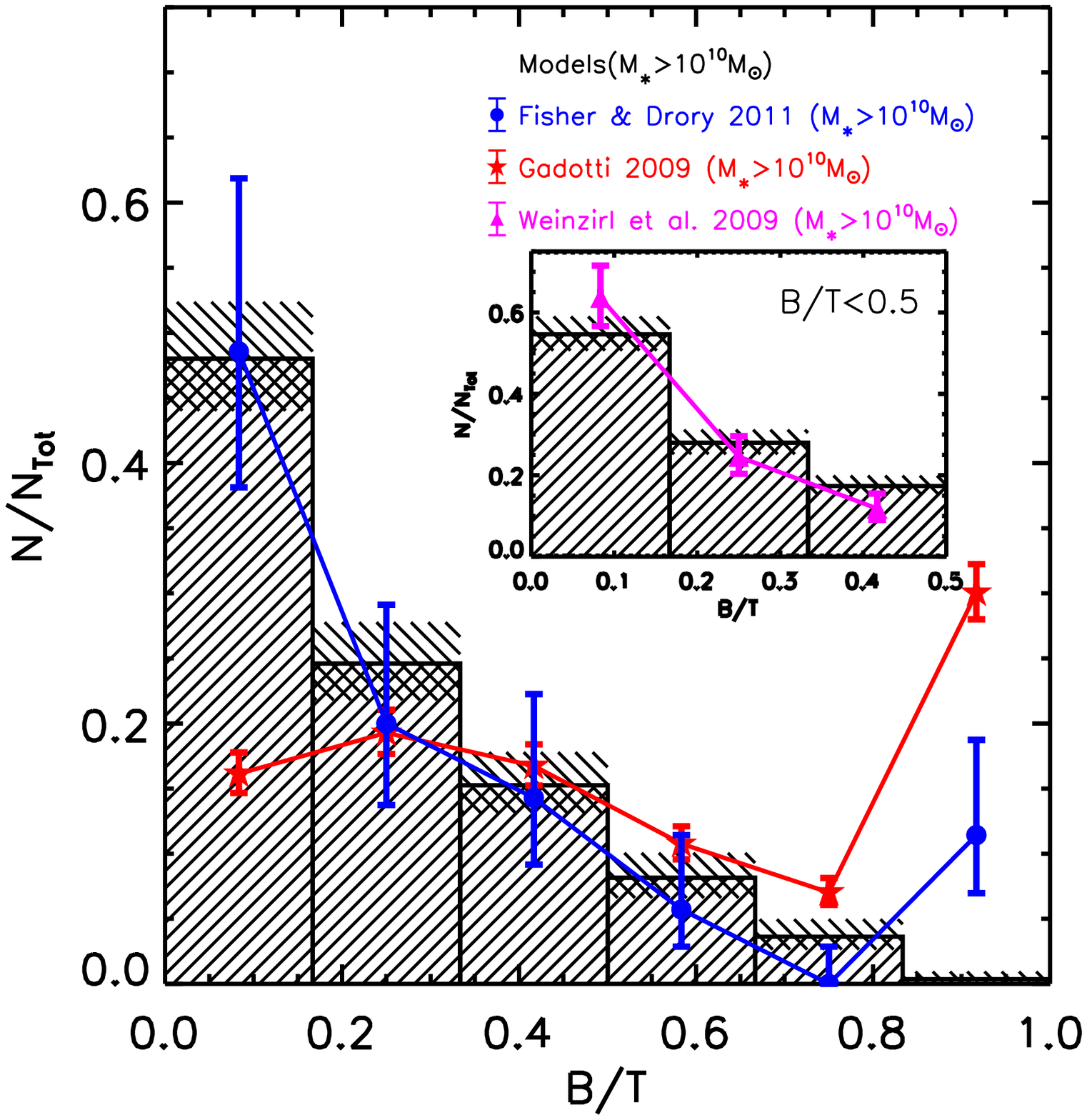}
\includegraphics[height=6.5cm,width=7.5cm,trim=0.95cm -0.05cm 0.8cm 0.85cm, clip=true]{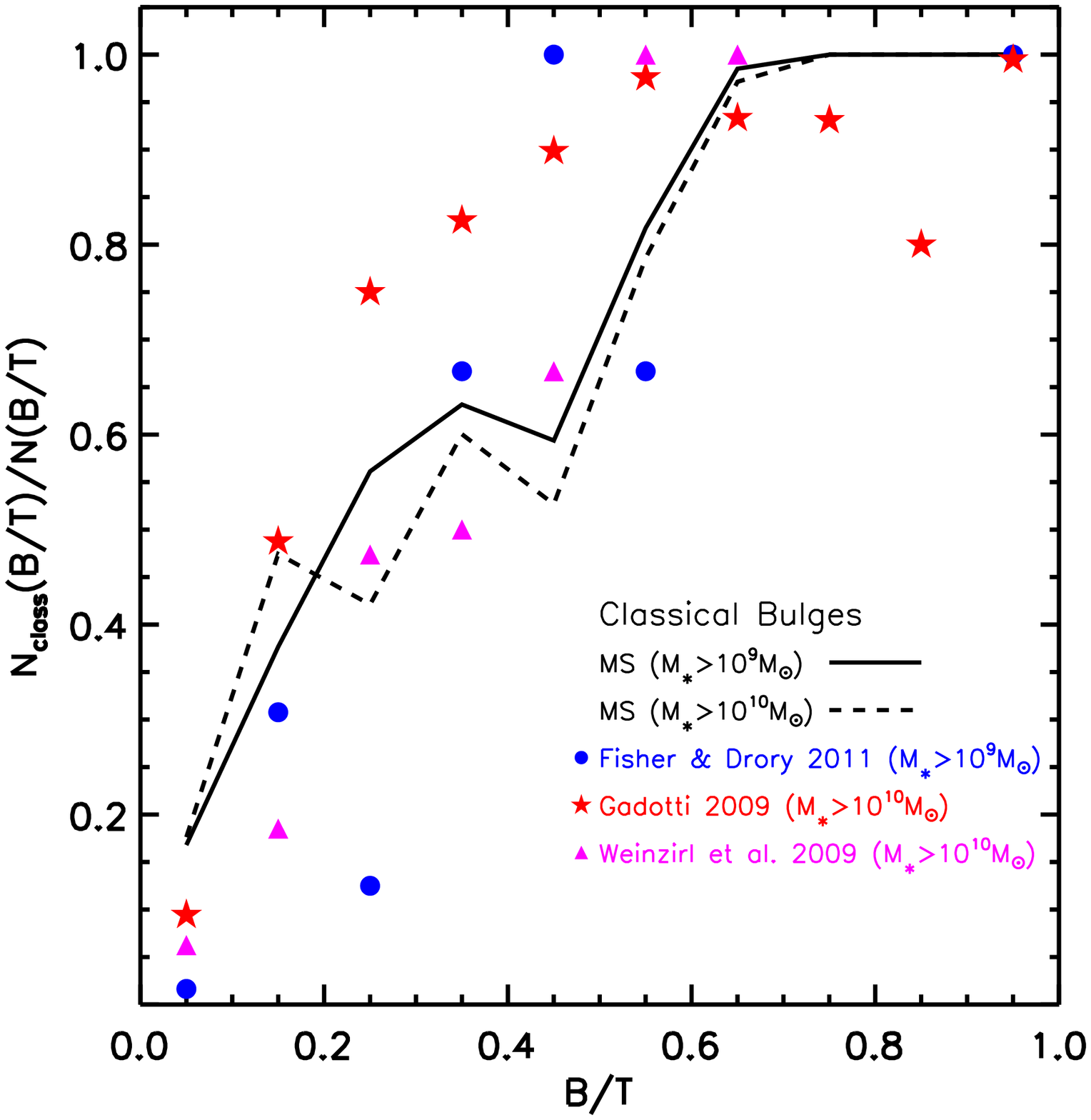}
\includegraphics[height=6.5cm,width=7.5cm,trim=1.8cm 0.5cm 0.85cm 1.0cm, clip=true]{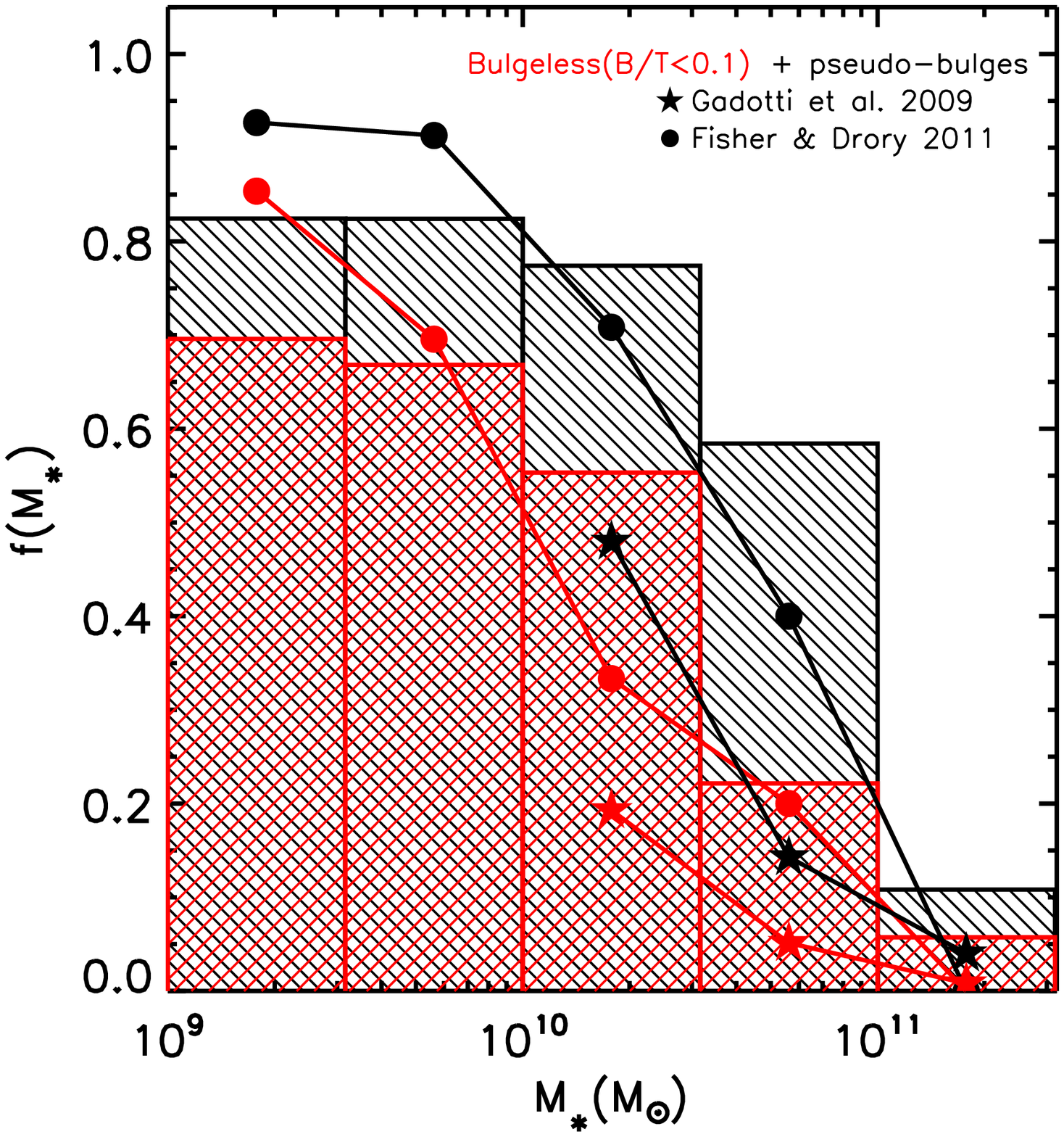}
\caption{{\it Top left}: $B/T$ distribution for galaxies with $M_{\ast}\geq10^{9}$M$_{\odot}$
 in the combined MS sample (black histograms) 
 and in the observational sample from FD11 (blue symbols). 
 {\it Top right}: The same as top left but for galaxies with $M_{\ast}\geq10^{10}$M$_{\odot}$; the galaxy samples from 
 G09 and W+09 have been added to the figure with red (star) and magenta (triangle) symbols, respectively.
 The latter is compared with the models only for $B/T<0.5$. Errors in the number counts are Poissonian and are marked with bars 
 for the observational data and with shaded regions for the model predictions.
 {\it Bottom left}: The combined MS sample with these different mass thresholds is divided in galaxies with a ``classical bulge'', 
 defined as those where the stars transferred from the primary contribute less than $50\%$ to the final bulge mass; the remaining 
 galaxies are called ``pseudo-bulges'' by extension. The panel shows the $B/T$ distribution
 for classical bulges with a solid (dashed) line for $M_{\ast}\geq10^{9}(10^{10})$M$_{\odot}$. The observational samples of G09 (red stars),
 FD11 (blue circles) and W+09 (magenta triangles) are also shown.
 {\it Bottom right}: Fraction of ``bulgeless'' galaxies as a function of \ms, with ``bulgeless'' defined as those galaxies with 
   $B/T<0.1$ (red), and these plus all pseudo-bulges with $B/T>0.1$ (black)}. The simulation data is shown
 with histograms and the observational data with stars and circles for G09 and FD11, respectively.
\label{histograms}
\end{figure*}

We first remark that the bulge/disc decomposition of observed galaxies and the characterization
of a spheroid as classical or pseudo-bulge are very difficult tasks
\citep[see e.g.,][]{Graham2001,MacArthur+2003,Allen+2006,Laurikainen+2007,
Fisher+2008,Tasca+2011,Simard-11}. There are only a few studies of bulge/disc decomposition applied to volume limited samples
that can be used to obtain fair statistics on the $B/T$ distribution as a function of \ms. We compare our predictions
with a couple of these studies in the following.

In the lower left panel of Fig.~\ref{BT_mass} we have included two
observational samples. The first (shown in red) is a volume-limited sample of 
$\sim1000$ local galaxies from the Sloan Digital Sky Survey (SDSS) with 
$M_{\ast}\geq10^{10}$M$_{\odot}$ taken from \citet{Gadotti2009} (hereafter G09). 
Since we are modeling central galaxies only, we have removed satellites 
from this sample by cross-correlating the original sample with the SDSS galaxy group catalog constructed 
by \citet{Yang_2007}. 
We find that $25\%$ of the original sample are satellites, $64\%$ centrals, and the rest are 
unidentified. We keep the latter two for comparison with 
our mock sample (we note however that we did not find a significant difference between the $B/T$ distribution of 
central/unidentified galaxies and that of satellites: the latter is just slightly more peaked at  $B/T\sim0.2-0.3$ than the former).
The median values of $B/T$ for the sample of central galaxies (solid red circles) increase with 
mass more steeply than our predictions, with median values larger by factors up to 1.5, although the scatter 
in the observational sample is very large. In particular, there is a significant fraction of observed galaxies with $B/T=1$. 

The second sample (shown in blue) is from FD11, which is
a volume-limited sample of galaxies within the local 11 Mpc volume. We have selected only 
those galaxies with $M_{\ast}\geq10^9$M$_{\odot}$ (a total of 99 galaxies).
Unfortunately, for this sample we do not know which galaxies are central/satellites.
At masses $9.8 < $log(\ms/\msun)$ < 10.8$,
the observed median values of $B/T$ are close to those of our predictions, 
well within the statistical scatter. However, at lower masses, 
the median $B/T$ ratios of our mock galaxies tend to be larger than those
of the FD11 sample. 

In the top panels of Fig.~\ref{histograms} we show the $B/T$ distribution for galaxies with 
\ms\ above a certain threshold ($10^{9}$M$_{\odot}$ and $10^{10}$M$_{\odot}$ 
for the left and right panels, respectively) for a combination of the MS-I and MS-II samples.
To obtain this combined MS sample, we used the latter up to $M_{h}=10^{12}\msun$ 
and the former for haloes with higher masses but with a number 
count renormalized to match the mass function of the MS-II sample in the mass bin $1-3.6\times10^{12}\msun,$ 
where both samples have neither a completeness nor a low number statistics problem. We have checked
that the results do not change significantly if the samples are taken separately in their respective 
mass range of completeness. The statistical error bars shown in the figure are Poissonian, employing the
definition given by \citet{Lukic_2007}:
$\sigma_{\pm}=\sqrt{N+1/4}\pm1/2$, where $N$ is the number of main haloes in a
given bin. We also show the observational samples from FD11 (blue), G09 (red) and 
W+09 (magenta). 
These panels show what was apparent in the left panels of Fig.~\ref{BT_mass}: the $B/T$
distribution strongly depends on the mass cut-off of the sample. 
For representative samples with a \ms\ threshold of $10^{9}$M$_{\odot}$, which are dominated
by low-mass galaxies, the fraction of galaxies with $B/T<0.1$ 
is 20\% higher than for samples with a threshold of $10^{10}$M$_{\odot}$.

The $B/T$ distribution of our mock galaxies roughly agrees with 
the one from the FD11 sample; although the latter
has a higher fraction of galaxies with $B/T<0.1$ and
slightly lower with $0.6\lesssim B/T \lesssim 0.9$.
We recall, however, that the FD11 sample
is very local and hence, massive galaxies (with typically high $B/T$ values) are undersampled.
These massive galaxies are represented better in the G09 sample, which shows a
flatter $B/T$ distribution than the one from FD11, and compared to our results, it has
a smaller fraction of galaxies with $B/T<0.2$ and a larger fraction of galaxies with $B/T>0.6$.  
The sample of W+09 has a morphological selection criterion that excludes elliptical
galaxies, and this is why we have restricted the comparison to $B/T\leq0.5$; for lower values of $B/T$, 
this sample has a similar distribution to the one from FD11, and thus, agrees with our predictions. 
If we assume that bulge-dominated galaxies ($B/T\ge 0.5$) correspond to E/S0 morphological types, 
then the fraction of such types predicted by our model is $\approx 16\%$ for $\ms>10^{10}$\msun. 
This is slightly lower than what was found observationally
in \citet{Baillard+2011}, where $\approx 20-25\%$ of galaxies are of E/S0 type
at $z\sim 0.015$ (for a sample with a completeness above $80\%$ for $\ms > 1.5\times 10^{10}$\msun).
Our fraction could be larger if one considers that some S0 galaxies actually have
$B/T<0.5$ \citep{Kormendy+2012}.

We note that our model is not able to produce galaxies with $B/T>0.9$. 
For the most massive galaxies, this could be because the primordial galaxies seeded at $z_{\rm seed}=3$ are
assumed to be discs, which implies that some fraction of these initial discs survives even after major mergers. 
In fact, several pieces of evidence suggest that the most massive galaxies (ellipticals with $\ms\gtrsim 5\times 10^{11}$ \msun) 
formed very early ($z>2$) and have not grown in mass by in situ SF since then \citep[e.g.,][]{Thomas+2005,Collins+2009}. 
If we were to assume that the most massive of our modelled galaxies where spheroids instead of discs at $z_{\rm seed}$, 
we would have a few $B/T\sim 1$ massive galaxies.

We also remark that in our scheme a disc may remain even after a merger with a very large mass ratio, contrary to 
some SAMs that assume that in these cases all the baryons end in a stellar spheroid \citep[e.g.,][]{Parry+2009,
deLucia+2011}. It should also be noted that (i) the semi-empirical $M_\ast(M_h)$ and $M_g(M_\ast)$ 
relations refer to average trends where additional dependencies on environment are not considered, 
and (ii) we do not follow the morphological evolution of satellite 
galaxies. According to the SAM results in \citet[][]{deLucia+2011}, the latter is not relevant to 
obtain $B/T>0.9$ galaxies. Instead, the possibility of disc conversion 
into a spheroid due to intrinsic disc instabilities \citep[see e.g.][]{Parry+2009}, which is something 
we did not take into account in our modelling, contributes to the difference with our predictions. 
While intrinsic secular evolution is a reasonable 
mechanism to produce pseudo-bulges in dynamically cold discs, it is a matter of debate if it could 
transform most of a massive stellar disc into an elliptical galaxy or destroy a small disc embedded into 
a dynamically hot spheroid.  

On the observational side, 
it is important to notice that in many cases of bulge/disc decomposition in elliptical galaxies, 
adding a small disc does not improve the statistical significance of a pure bulge model, and a value of 
$B/T=1$ is simply assigned. However, several studies provide evidence to argue that most ellipticals contain small 
discs \citep[e.g.][]{Ferrarese-1994,Lauer-2005}.

\begin{figure}
\centering
\includegraphics[height=6.5cm,width=8.5cm]{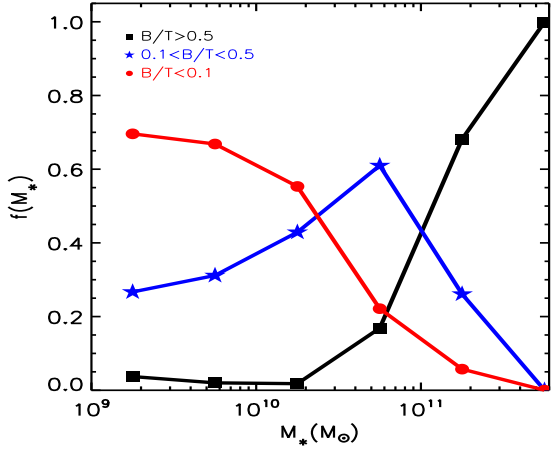}
\caption{Predictions of the relative fraction of galaxies as a function of \ms\ 
  according to the final bulge-to-total mass ratio:
  $B/T<0.1$ (red circles), $0.1<B/T\le 0.5$ (blue stars) and $B/T>0.5$ (black squares).}
\label{BTfractions}
\end{figure}

Fig. \ref{BTfractions} is another way of showing the predictions of our model. 
Here we plot the fractions of galaxies in different \ms\ bins divided
according to their $B/T$ ratios: $B/T<0.1$ (filled circles, red line), $0.1\le B/T<0.5$ (stars, blue line), and
$B/T\ge 0.5$ (squares, black line).  Roughly, these three groups can be associated to irregulars (types
later than Sc),  intermediate disc-dominated (Sa to Sc types), and bulge-dominated (E/S0) galaxies, respectively. 
At $\ms<10^{10}$\msun,  $\approx 65\%$ of our model galaxies have $B/T<0.1$, at $\ms\approx 3-8\times10^{10}$ 
\msun\ more than 50\% have $0.1\le B/T<0.5$, while at $\ms>10^{11}$ \msun,
galaxies with $B/T\ge 0.5$ dominate. These predictions are roughly consistent with estimates of the local
morphological mix. 
The predicted mass fractions of stars contained in discs ($\ms>10^9$ \msun) is 57\% and the
rest is in spheroids, consistent with observations. The fraction of stars in galaxies with 
$B/T\le 0.1$ is $\approx 15\%$, while the fraction of stars (both in the disc and the spheroid)
in galaxies with $B/T>0.5$ (spheroid-dominated) is $\approx 58\%$.

\subsubsection{Comparison with observations: bulge composition}

We study the fraction of galaxies whose bulges were assembled mostly by stars from 
the secondaries (contributing to the bulge mass fraction by more than $50\%$). 
This fraction as a function of $B/T$ is shown in the bottom left panel of Fig.~\ref{histograms}
(solid and dashed lines for  $\ms>10^{9}$M$_{\odot}$ and $>10^{10}$M$_{\odot}$, respectively).
Because this channel of bulge growth is dominant when stellar major mergers dominate 
the bulge mass assembly (see right panels of Fig.~\ref{BT_mass}), 
we will nominally refer to these bulges as classical-like (CL). 
On the contrary, those bulges where more than 50\% of their stars come from the primary will be nominally
defined as pseudo-like (PL) bulges. Notice that this division is well motivated: classical 
(pseudo) bulges are thought to be formed by major mergers (disc secular evolution/minor mergers). 
Thus, one expects that classical (pseudo) bulges are dominated by stars acquired from the 
secondaries (primary discs). Moreover, the kinematics of the bulge stars, often regarded as one of the key
discriminants between pseudo and classical bulges, are expected to be consistent with this division: 
(i) bulges whose stars come predominantly from the secondaries will have more random orbital orientations 
(not necessarily aligned with the disc) and a bias towards radial anisotropy, (ii) bulges where 
most of the stars were scattered by instabilities resonantly excited by mergers
(or formed ``in situ'' in the starburst) will have the bulge 
preferentially aligned with the disc, with more rotation, and tangentially biased orbits.  

\emph{We find that most of the bulges in our mock galaxies 
with $B/T\gtrsim 0.25$ are CL, while the ones of small galaxies are typically PL}. For 
$B/T<0.1$, only $\lesssim25\%$ of the bulges are CL. The red stars, blue 
circles and magenta triangles on this figure correspond to the observational samples of G09, 
FD11 and W+09, respectively, where the bulge class is determined
from the light profile (photometry). Our prediction roughly agrees with these observational samples, 
although it slightly under(over)-predicts  the abundance of classical bulges in the 
intermediate (low) $B/T$ range. 

In the bottom right panel of Fig.~\ref{histograms} we show the fraction 
of galaxies with $B/T<0.1$ (which we call ``bulgeless'', red histogram) as a function of \ms. For the
black histogram we add to this fraction those galaxies with $B/T\ge0.1$ that have PL bulges. 
The circles and stars correspond to the respective fractions for the samples of
FD11 and G09, respectively. 
The deficit of CL bulges for intermediate $B/T$ values shown in the lower left panel appears 
in this figure as an overprediction of PL bulges for galaxies
with $M_{\ast}\approx 5\times10^{10}$M$_{\odot}$ by a factor of $\sim1.5$.

Our model seems to underpredict the fraction of bulgeless 
galaxies by less than $15\%$ at $\ms<10^{10}$ \msun.  
At larger masses, the agreement is good although there is an excess of bulgeless galaxies, which
actually grows to more than a factor of $\sim 2$, once we compare with the G09 sample; 
compared to the W+09 sample however ($\ms>10^{10}$ \msun~disc galaxies),
our predicted fraction is slightly lower. 
Thus, it seems that the potential problem of too few bulgeless galaxies is related only to galaxies with 
$\ms<10^{10}$ \msun.  We recall that the FD11 sample contains both central and satellite 
galaxies, while our results refer only to the former. Because centrals have a higher probability
to suffer mergers than satellites, with the difference being larger for smaller \ms, we expect that bulgeless 
galaxies are more frequent in low-mass satellites than in centrals. Indeed, using the \citet{Wang+2008} SAM and the 
H09a procedure to estimate bulge growth, \citet{Fontanot+2011} found that, at a given mass, low-mass satellites  
are more likely to be bulgeless than centrals. Unfortunately, we do not know the fraction of 
satellites in the FD11 sample, but if we could exclude it, it is likely that the bulgeless fraction 
would be lower than the one shown in the bottom right panel of Fig. \ref{histograms}.

The predicted fraction of bulgeless galaxies could also decrease if one takes into account the effect of 
intrinsic disc instabilities, although \citet{Fontanot+2011} found that this is only a minor effect for 
low-mass central galaxies. Despite that this effect is more relevant for massive galaxies, most of these 
already have large $B/T$ ($>0.1$) ratios due to major mergers. 

\section{Summary and Conclusions}\label{sec_concl}

We present a scenario of bulge growth in {\it central} galaxies based on: \textit{(i)} the mass 
aggregation and merger histories of \lcdm\ distinct haloes randomly selected from both Millennium simulations 
(\S 2), \textit{(ii)} the use of $z$-dependent empirical relations, \ms(\mh) and \mg(\ms),  
to seed galaxies into the growing (sub)haloes (\S 2.3, Appendix A1),
and  \textit{(iii)} the implementation of a physically motivated model 
to calculate the bulge stellar mass after a galaxy-galaxy merger (Appendix A2). 
By construction, the stellar mass assembly of our mock galaxies follows on average the
general downsizing trends inferred from observations (see \citealt{FA10}).

In our approach, bulges grow through galaxy mergers 
characterized by the dynamical mass ratios of the merging galaxies
at the coalescence time. The latter is estimated following the orbit of the subhalo until 
its number of particles falls below the resolution of the simulations, and later 
applying a tested approximation of the dynamical friction time (Eq.~\ref{merge_time}).
The stellar and gas masses of the secondaries at the moment of coalescence 
are either assumed to be the same they had when 
their haloes became subhaloes 
(extreme satellite quenching) or their further 
evolution until coalescence is estimated using a simple model of SF and SN-driven 
outflows (Appendix \ref{semi_app_3}), under the assumption that there is no further gas accretion. 
The bulge stellar mass increases after a merger by three contributions:
(1) all the stars from the secondary, (2) a fraction of stars transferred from the primary disc, 
and (3) stars formed in a central starburst triggered by the
condensation of cold gas from both merging galaxies. 
While the first contribution is expected to be related to 
a classical bulge, the other two are associated to
a pseudobulge, i.e. a bulge that shares properties with the disc. 

Our goal has been to revisit and explore in detail the predictive power 
of the \lcdm~paradigm to explain, through the halo-halo merger histories mapped in a non-trivial way to 
galaxy-galaxy merger histories, the properties and demographics of the bulges of central galaxies
with $\ms\geq10^{9}$ \msun. We highlight the following results and conclusions:

$\bullet$ Mock galaxies with $\ms\lesssim 2\times 10^{10}$ \msun\ have on average $B/T<0.2$, 
while for more massive galaxies, $B/T$ increases rapidly with \ms. 
This is mainly because the stellar major merger rate history increases with mass.
For $\ms<10^{10}$ \msun, nearly 60\% of galaxies have $B/T<0.1$, while for $\ms>10^{11}$ \msun\
galaxies with $B/T>0.5$ dominate.
The driver of bulge growth in massive/high-$B/T$ galaxies are major/intermediate 
($\mustar>0.1$) stellar mergers. Instead, for low-mass/low-$B/T$ galaxies, bulges grow
mainly through minor/minuscule stellar mergers ($\mustar<0.1$), although the corresponding 
dynamical merger ratios are larger, $\mu_{\rm eff} > \mustar$. 
The latter implies that the bulges of~$\lesssim10^{11}$ \msun~galaxies have grown mostly from stars of the 
primary disc. 

$\bullet$ Bulges are composite, growing in several episodes through concomitant channels of 
stellar mass acquisition: from the secondaries, from the primary disc, and from starbursts. 
The first (second) of these contributions increases (decreases) monotonically with \ms.
For $\ms\approx 10^9, 3\times 10^{10}$, and  $10^{11}$ \msun, approximately 60\%, 55\%, and 45\% of the bulge 
mass was assembled by stars from the primary, respectively. For $\ms\gtrsim10^{11}$ \msun, 
the stars acquired from the secondaries contribute $>50\%$ to the bulge mass. 
Bulge stars formed by central starbursts contribute $<5\%$ to the
bulge mass on average. Although intrinsic disc instabilities were not taken into account as a mechanism of
(pseudo)bulge formation, the merger-induced disc instabilities produce a similar effect
regarding bulge mass acquisition. Since this mechanism dominates at low \ms, the stars of the 
bulges of small galaxies are predicted to have similar properties, including their kinematics, than the 
stars of their discs.
We can nominally consider that when stars from the secondaries contribute more (less) than
$50\%$ to the final bulge mass, then the bulge is classical-like (pseudo-like). The fraction of 
classical-like bulges is $>50\%$ for galaxies
with $B/T\gtrsim 0.25$. Galaxies with smaller $B/T$ values have mostly pseudo-like bulges. 

$\bullet$ The evolution of the satellites from the time of infall until final coalescence influences
the $B/T$ ratio of the central galaxy and significantly affects the ratio of secondaries-to-primary 
stars in the remnant bulge. In the extreme case of total SF quenching in the satellites, low-mass galaxies end up 
with smaller $B/T$ ratios than in the case where SF/feedback is allowed. 
In the former case, the contribution of stars from the secondaries to the growth of the 
bulge falls to $\lesssim10\%$, resulting in a negligible fraction of classical-like bulges for galaxies 
with $\ms\lesssim 10^{11}$ \msun, which is in conflict with observations. On the contrary, 
if satellites keep and transform their gas into stars 
too efficiently, then the mergers would be very destructive and the central galaxies would end up with
$B/T$ ratios that are too high. An intermediate case between the previous two extreme cases is
the standard SAM we have used to follow SF/feedback 
in the satellites (see Appendix \ref{semi_app_3}); it assumes that their hot gas is completely stripped
as they merge with the host\footnote{If environmental processes such as ram pressure stripping not only remove
all the hot gas from the satellites but also their cold gas discs, then star formation will be further suppressed
and it would be more difficult to reproduce the observations.}. For this case, it turns out that on average,
the stellar mass of a satellite of a given mass grows in a similar way
as that of a central galaxy of the same mass, with the latter being determined by the evolution of the \ms(\mh)\ 
relation. Since this case is quite successful, as we summarize below, the latter finding is also
very useful as a reference to infer merger rates from observations based on pair samples 
at relatively large separations.

{\it Are our results in agreement with observations of bulge/disc demographics?}
Firstly, we checked that the
predicted stellar merger rates as a function of mass and $z$ are actually consistent with 
observational measurements (Fig. \ref{frac_merger}).
Then, we compared our results with the observational samples of FD11, G09, and W+09
regarding the $B/T$ dependence on \ms\ (Fig. \ref{BT_mass}), the $B/T$ distribution and the fractions of galaxies 
with a given $B/T$ and \ms\ (including the fractions of classical- and pseudo-like bulges; Fig. \ref{histograms}). 
Despite the large observational uncertainties, the overall consistency between our \lcdm-based predictions 
and observations is remarkable. In particular, it is quite relevant that a large fraction of the pseudo-like 
bulge population can be explained in a merger-induced scenario without introducing pure intrinsic secular instabilities
on the disc. We also find an agreement in the fractions of galaxies within a given $B/T$ interval, which can be roughly 
associated to different morphological types 
(the morphological mix, Fig. \ref{BTfractions}), as well as in the fractions of stars in the disc and 
bulge components. While overall our prediction for the $B/T$ distribution is consistent 
with observations, in detail we have detected some potential disagreements: 

-There are no mock galaxies with $B/T\sim 1$, 
while observers assign $B/T=1$ to a small fraction of (discless) galaxies. We argue that this apparent
disagreement rather than a failure of the model could be a prediction: even the giant
ellipticals should have small hidden discs.

-At the mass range $10^9<\ms/\msun<10^{10}$, there are $\approx 15\%$ fewer mock galaxies
with $B/T<0.1$ than observed in the local 11 Mpc volume sample of FD11. The latter, however, is likely to 
contain a non-negligible number of satellites that increase the fraction of bulgeless galaxies 
(satellites are expected to have smaller $B/T$ ratios than centrals of the same mass). 
On the other hand, the fraction of low-mass $B/T<0.1$ 
mock galaxies can easily be increased if the low-mass end of the $\ms(\mh,z)$ relation is such 
that at higher redshifts the values of \ms\ for a given \mh\ are smaller than those used here
(see Appendix B).

We conclude that the implementation of a reasonable semi-empirical model of galaxy occupation into
growing \lcdm\ haloes is able to predict the present-day demographics and mass dependence of the 
galaxy $B/T$ ratios. At the basis of our results are 
the \lcdm\ halo merger rates, the merger-driven bulge growth model,
and the way galaxies assemble their gas and stars, with a 
downsizing trend and a bell shaped (low) stellar formation efficiency. 

Although these results might be interpreted as a positive test of the \lcdm\ model, we remark that 
we have only included merger-driven mechanisms of bulge growth.
Discs instabilities such as bars and spiral arms also lead to secular processes
of pseudobulge growth, in particular, the "peanut-shaped"/boxy bulges are thought to be 
a byproduct of bar formation \cite[e.g.][]{Kui_95}. There are other two possible mechanisms of 
bulge formation: (i) fragmentation of the gas-rich disc into clumps that migrate towards the center forming a 
large bulge at high redshifts \cite[][although stellar feedback might suppress this mechanism, see 
\citealp{Hopkins11}]{Dekel_09}, and (ii) 
misalignment of the angular momentum of the disc with that of the newly-accreted gas \citep{Sales-11}.
All these non-merger processes would probably increase
the $B/T$ ratios of our mock galaxies, decreasing the population of bulgeless galaxies. Since our predictions are already
in marginal agreement with observations of low-mass galaxies, the addition of these mechanisms might pose
a challenge for the \lcdm\ model. 

\section*{Acknowledgments}

We are grateful to the reviewer, Philip Hopkins, for a very thorough and constructive report.
We would like to thank Simon D. M. White, Michael L. Balogh, Dimitri Gadotti, Francesco Shankar and Dave
Wilman for interesting comments and suggestions. JZ is supported by the University of Waterloo and the Perimeter
Institute for Theoretical Physics. Research at Perimeter Institute is supported by the Government of Canada through
Industry Canada and by the Province of Ontario through the Ministry of Research \& Innovation. 
JZ acknowledges financial support by the Joint Postdoctoral Program in Astrophysical Cosmology
of the Max Planck Institute for Astrophysics and the Shanghai Astronomical Observatory, and from a CITA National
Fellowship. VA acknowledges PAPIIT-UNAM grant IN114509. MBK acknowledges support from the Southern California 
Center for Galaxy Evolution, a multicampus research program funded by the University of California Office of Research.

\bigskip

\bigskip

\appendix

\section{Semi-empirical model of galaxy occupation}\label{semi_app}

\subsection{Stellar and gas mass relations}
\label{semi_app_1}

Along the evolution of each halo we assign 
stellar masses by using semi-empirically determined average stellar-to-halo mass relations 
at different epochs. By means of the abundance matching technique, \citet{Behroozi-Conroy-Wechsler-10} 
determined the \ms--\mh\ relations for central galaxies in the $0< z\lesssim1$ range and then 
extended these relations to $1\lesssim z<4$. These authors presented two different fitting formulae
to their results, one for each redshift range. Because these formulae are disjoint at $z\sim 1$,
we use instead the function presented in \citet{FA10}, who modified them in order to obtain a continuous 
function from $z=0$ to $z=4$. 

Each galaxy is assigned a certain cold gas mass by drawing values from a Gaussian distribution 
with a mean and standard deviation according to the stellar-to-gas mass 
function $M_g(M_\ast,z)$ (fitted to observations) given by \citet{Stewart-09}. 
The extrapolation of this function to very low masses can produce unrealistic gas fractions. 
To avoid such cases, \citet{Stewart-09} proposed to assign an upper
limit to the gas fraction given by $f_{\rm lim}(z)=M_g/M_h\vert_{M_\ast=3\times10^8{\rm M_{\odot}},z}$ such that $M_g(z,M_\ast)\leq
f_{\rm lim}(z)M_h(z,M_\ast)$. The value of $M_\ast=3\times10^8M_{\odot}$ was
probably chosen because is the minimum mass for which a measurement of
the cold gas fraction was inferred in the work of \citet{Stewart-09}. Given the
high degree of uncertainty in the values of $f_{\rm gas}$ for low-mass galaxies,
increasingly higher for higher $z$, we opt to simply set $f_{\rm gas, lim}\leq100$, which is the
maximum observed value reported in \citet{Stewart-09}.
 
As a halo grows, the stellar and cold gas masses of its
central galaxy are given by the aforementioned relations (with the new stellar mass added to
the disc component), except during a period
of time after final coalescence in a galaxy-galaxy merger. In this case, 
there is an ``instantaneous'' increase of the stellar mass of the remnant that is naturally 
higher than the value given by the mean $M_\ast(M_h)$ relation. In such cases, the
halo-to-stellar mass relation is only restored in the future once the halo has grown
sufficiently. For some cases, this moment has not happened yet by
$z=0$ and thus, the final sample of mock galaxies has slightly higher stellar masses than
predicted by the $M_\ast(M_h,z=0)$ relation.  We have verified that the shape of the
stellar mass function that we recover at $z=0$ is roughly in agreements with that of the 
observed relation by comparing to the best Schechter fit reported in \citet{Panter_2007}. 
Overall, the values are always within a factor of $\sim2-3$ with a spread that is larger for 
higher masses (because in average, more massive galaxies have more recent major mergers). 

The redshift we choose to seed the galaxy population is $z_{\rm seed}=3$ since for higher
redshifts, the stellar and cold gas mass assignment becomes highly uncertain. Galaxies are
seeded as pure discs at $z_{\rm seed}$, thus, we are effectively neglecting their previous 
morphological evolution. This is partially justified
observationally since 
the fraction of disc-dominated galaxies gets larger with higher redshift. In particular, at
$z\sim1$, \citet{Pannella-09} report a transition stellar mass of $10^{11}$M$_{\odot}$ below
which disc-dominated galaxies dominate the galaxy population. Similarly,
from extrapolating the results of \citet{Oesch-10} to $z\sim1$, we estimate that $\sim90\%$ 
of all galaxies with $M_{\ast}<10^{11}$M$_{\odot}$ are disc-dominated ($B/T<0.5$),
whereas for larger masses, the mix is approximately $75\%$ and $25\%$ for
disc- and bulge-dominated galaxies, respectively. For $z=3$, the mix would
be even more in-balanced towards mostly disc-dominated systems. 

\subsection{Channels of merger-driven bulge growth}
\label{semi_app_2}

\subsubsection{ Transfer of stars from the primary and secondary into the bulge.} 

In its orbit through the
primary, the secondary looses energy and angular momentum sinking to the center due to dynamical
friction. In the last stages of coalescence, the surviving mass of the secondary $M_2$ (dark
matter + baryons) collides with the central region of the primary, of mass $M_1$. During this process,
the collisionless components of both systems are subjected to rapid changes
of the gravitational potential that broaden their energy distributions leading towards an equilibrium
state. This violent relaxation process drives the stars originally 
rotating in discs towards random orbits forming a spheroidal remnant. 
The stars in the primary disc affected by the dynamical action of the secondary are expected to 
be those within a radius enclosing the mass $\sim M_2$.
The stars at larger radii in the disc are also perturbed but likely
they are re-arranged into final configurations that are not far from the original ones 
(for instance, one of the effects of the merger is to vertically heat the galactic disc). It is then reasonable 
to expect that approximately a fraction: 
\begin{equation}\label{violent_rel}
f_{\rm relaxed}^p=\frac{m_{\ast,{\rm disc_{rel}}}^p}{m_{\ast,{\rm disc}}^p}\approx\mu_{\rm eff}\equiv\frac{M_2}{M_1}
\end{equation}
of stars in the primary plus all the stars in the secondary will pass through violent relaxation and be 
transferred into the primary bulge. Due to the highly non-linear nature of mergers, they are better studied with
numerical simulations, like those carried out in H09a. They seem to confirm the approximation described before, 
although a more detailed -but uncertain- accounting of the efficiency of violent relaxation as a function of radius suggests a
slightly non-linear dependence on $\mu_{\rm eff}$. The correction factor to Eq.~(\ref{violent_rel}) 
suggested in \citet{Hopkins-09b} is $2(1 + \mu_{\rm eff}^{-a})^{-1}$, with $a=0.3-0.6$. We adopt
this correction and use $a=0.3$ for our fiducial model, noting that the values of $f_{\rm relaxed}^p$ 
are thus smaller than those given by Eq.~\ref{violent_rel}, making 
violent relaxation less efficient in the primary disc and slightly decreasing the final values of $B/T$,
particularly at low masses.

\subsubsection{Starburst from the transport of cold gas into the centre of the primary.} 

A fraction $f_{\rm burst}$ of the cold gas in the galaxies looses 
angular momentum because the gravitational
interaction during final coalescence generates a non-axisymmetric response in the galactic discs
that morphologically resembles a bar. The resulting stellar and gaseous bars are however out of phase
because gas is collisional and stars are not. Because of this, the stellar bar torques the gas bar draining
its angular momentum. In this way, the cold gas is effectively removed from the original discs 
and transformed into stars, during a starburst, in the bulge of the remnant.
This process is efficient within a region internal to a critical radius $r_{\rm crit}$
and thus, the fraction $f_{\rm burst}$ (assuming that the discs have exponential surface density
profiles) is given by:
\begin{equation}\label{m_burst}
  f_{\rm burst}=\frac{m_{\rm burst}}{m_{\rm cold}}=A\left[1-(1+r_{\rm crit}/r_{d})e^{(-r_{\rm crit}/r_d)}\right]
\end{equation}
where $r_d$ is an effective gaseous disc scaling radius, and $A$ is a normalization factor. 
The ratio of the critical to effective scale radius
depends on the merger mass ratio and relative orientation and orbit of the progenitors, as
well as their stellar and gaseous content. A parameterization of this ratio is given 
in Eq. A2 of \citet{Hopkins-09b}.
We find that using Eq.(\ref{m_burst}) with
$A\sim1/0.26$ gives a reasonable match to the simulation results
shown in Fig. 7 of H09a. We note however that  the precise
value of the normalization is not very relevant for our results (unless
$A\gg1$). This is because the cold gas mass that participates in the starburst is
usually subdominant compared to the stellar mass that is violently relaxed in
the primary and secondary.

The total mass of the stellar disc and bulge of the remnant after a merger are then given by:
\begin{eqnarray}\label{masses}
  m_{\ast,{\rm disc}}^r&=&m_{\ast,{\rm disc}}^p(1-f_{\rm relaxed}^p)\nonumber\\
  m_{\ast,{\rm bulge}}^r&=&m_{\ast,{\rm bulge}}^p+m_{\ast,{\rm disc}}^pf_{{\rm relaxed}}^p+m_{\ast}^s\nonumber\\
  &+&f_{\rm burst}^pm_{\rm cold}^p+f_{\rm burst}^sm_{\rm cold}^s
\end{eqnarray}
where the superscripts $p$ and $s$ refer to the primary and secondary, and all
the quantities on the right are defined just before coalescence; it is then necessary to estimate the dark matter and
baryonic mass of the galaxies at this time. 

For the primary, the effective dark matter mass is that of the most bound material 
just prior to coalescence, which is well approximated by
replacing the real density distribution by a NFW radial density profile and
computing the mass interior to a radius 
$r_s=r_{\rm vir}/c$, where $c$ is the halo concentration (the c(\mh,$z$) relation 
of \citet{Gao-08} is used). At this stage, its baryon content is simple given by
the $M_\ast(M_h,z)$ and $M_g(M_\ast,z)$ relations defined earlier.

The effective dark matter mass of the secondary is computed from the
properties its halo had when it was about to enter the virial radius of the
host. Due to tidal striping, most of its dark matter mass 
will be lost before final coalescence, but the most bound material should survive.
Because of this process, the subhalo mass in the final stages of the
merger is an unreliable tracer of the potential well that shaped the
properties of the satellite galaxy. For this reason, the empirical
$M_\ast(M_h)$ and $M_g(M_\ast)$ relations only apply to the secondary before 
it enters the virial radius of the main halo, and not afterwards as it
spirals inwards. {\it Modeling the evolution of the stellar and gaseous components
of the secondary is an important element of the semi-empirical model we have used
in this paper}; we discuss it below.

\subsection{Evolution of satellite galaxies}
\label{semi_app_3}

A first simplistic approach would be to assume that the amount of
cold gas and stars in the satellite just before coalescence is the same as in
the moment the merger started (the results in the first row of Fig.~\ref{BT_mass} are obtained
using this approach). A more physically-motivated treatment, used in our fiducial model, 
is to assume that there is no further gas accretion
(due to environmental processes such as starvation, ram pressure and tidal striping) 
but the satellite suffers SF and feedback processes. For the former, we assume
the Kennicutt-Schmidt law \citep{Kennicutt-98} for gas surface
densities $\Sigma_g$ above a critical density $\Sigma_c=10{\rm M_{\odot} pc^{-2}}$:
\begin{equation}\label{SFR}
{\dot \Sigma}_\ast=0.25~{\rm M_{\odot} pc^{-2}
  Gyr^{-1}}\left(\frac{\Sigma_g}{{\rm M_{\odot} pc^{-2}}}\right)^{1.4}
\end{equation}
Initially, the disc has an an exponential gas surface density with an effective
scale length given by $2$
times the effective radius of the disc, which we take to be given by the 
following empirically motivated relation\footnote{Hopkins, P., private communication, based on the
results of \citet{Shen-03}.}:
\begin{equation}\label{star_to_size}
r_{\rm eff}=5.28{\rm kpc}~(1+z)^{-0.6}\left(\frac{M_\ast}{10^{10}{\rm M_{\odot}}}\right)^{0.25} 
\end{equation}
We then apply Eq. (\ref{SFR}) 
for azimuthally averaged values of $\Sigma_g$ in rings
to obtain the mass of new stars formed in a time interval
$dt$ as a function of galactocentric radius $R$:
\begin{equation}\label{new_stars}
dM_\ast(R)=(1-\mathcal{R})\pi A_a(R) {\dot \Sigma}_\ast(R) dt 
\end{equation}
where $A_a(R)$ is the area of the annulus at radius $R$, and $\mathcal{R}=0.35$ is the fraction of the mass in 
stars that is instantaneously returned to the cold gas. This formula is reasonable 
as long as $dt$ is smaller than the overall time scale of SF. 

For our fiducial model, we also adopt a recipe for a SN feedback-driven outflow
following the energy-driven wind case of
\citet{Dutton-vandenBosch-09}. The amount of cold gas ejected in each annulus is then given by:
\begin{equation}\label{feedback}
dM_{\rm eject}(R)=\frac{2\epsilon_{\rm EFB}E_{\rm SN}\eta_{\rm SN}}{V_{\rm esc}^2(R)}dM_\ast(R)
\end{equation}
where $E_{\rm SN}=10^{51}$ergs is the energy of one SN,
$\eta_{\rm SN}=8.3\times10^{-3}$ is the number of SN produced per solar
mass of stars, and $V_{\rm esc}(R)$ is the escape velocity at radius $R$ given by
the gravitational potential of baryons and dark matter.
The kinetic energy of the wind is assumed to be a fraction $\epsilon_{EFB}=0.16$ of the SN energy. 

Using Eqs.~(\ref{SFR}-\ref{feedback}) we estimate the gas and stellar content of
the satellite at the time of coalescence (recall
that the merging time-scale is described in section \ref{times_section}). 

Since our model is not aimed to follow the bulge growth of satellites,
we cannot assign $B/T$ values to them. This impacts the amount of cold gas in the
secondary that participates in the central starburst during a merger. By testing extreme $B/T$ values for the
satellite, from pure discs to pure bulges, we find that, in most cases, the morphology of the
satellite is not a significant factor in the final morphology of the
remnant. This is mainly because the mass that participates in the
starburst is usually lower that the stellar mass that is violently
relaxed in the merging galaxies (see also Fig. 14 of \citealt{Hopkins-10a}).

\section{Variations of the semi-empirical model}\label{variation}

We study now, in an almost qualitative way, the variations to some of the relevant 
ingredients of the semi-empirical model of galaxy occupation, 
concentrating on the statistical changes to the morphological mix of galaxies at $z=0$.

{\it Changes in $M_{\ast}(M_h,z)$. } 
A substantial change in the evolution of the $M_{\ast}(M_h,z)$ relation has consequences 
for the efficiency of bulge growth through mergers. We explored this using 
the relation reported in \citet{Yang+2011} for case 12 in their Table 3. 
At $z=0$, the difference between this formula and that of 
\citet{FA10} is not significant, except at the low-mass end where the former 
predicts lower values of $M_{\ast}$ for a fixed halo mass. At high redshift however, both formulae 
differ significantly in the low-mass end, with the \citet{Yang+2011}
formula predicting lower stellar masses for a given halo mass. This means that secondaries will be less destructive 
for central galaxies. The net effect is an overall reduction in the values of $B/T$, except for the 
most massive galaxies, whose morphologies are typically shaped by  
massive secondaries. Thus, by using a $M_{\ast}(M_h,z)$ relation with systematically lower
stellar masses in the low-mass end, the morphological mix changes towards a  
more extreme dominion of bulgeless
galaxies in this mass regime. For example, for the case analysed, we find that at $M_{\ast}=10^{10} \msun$, $\sim85\%$ of
galaxies have $B/T\le 0.1$, which is $\sim1.7$ times more than in our fiducial case, whereas there is essentially the same 
morphological mix for $M_{\ast}\geq10^{11}\msun$.

{\it Gas fraction limit. } 
The $f_{\rm gas}(M_{\ast},z)$ empirical relation from \citet{Stewart-09} that we have used 
to assign a gas fraction to our mock galaxies can be trusted up to $z=2$ down to 
$M_\ast\sim3\times10^8$M$_{\odot}$ (see Fig. 2 of \citealt{Stewart-09}).
For higher redshifts and lower masses, there is essentially no data to constrain this relation. 
In our fiducial model, a higher value of $f_{\rm gas}$ at the start of the merger implies that the 
secondary will have more material available for SF as it spirals inwards towards the primary, 
resulting in a more destructive merger. We investigate the impact of 
a limiting value of $f_{\rm gas}$ different to the one used in the 
fiducial model ($f_{\rm gas,lim}=100$). This alternative 
value is given by the cosmic baryon fraction: 
$M_{gas}+M_{\ast}\leq f_bM_h$ with $f_b=0.17$. In this case, all of our mock galaxies end up being more massive at $z=0$, the change
being significant for the least massive galaxies and negligible for the most massive ones. The values of $B/T$ are systematically
higher, specially in the low-mass end, which produces a morphological mix that is deficient in bulgeless galaxies. For instance,
at $10^9\msun$, the models predict that only $20\%$ of the galaxies have $B/T<0.1$, while for
$M_{\ast}\geq10^{11}\msun$, there is essentially no change compared to our fiducial model.

{\it SF and feedback recipes for the secondary. } The satellite evolution that we have chosen in 
our fiducial model is based on the assumption that the Kennicutt-Schmidt law is valid 
at all redshifts.It is possible however, that this law changes with $z$,  
being steeper and with lower normalization at higher $z$ 
(e.g., see Fig. 3 of \citealt{Gnedin_2010}).
A lower SF rate at higher $z$ will result in satellites with higher gas fraction at the moment of 
final coalescence and thus, in mergers that are less destructive. 
We find that within the range of values of the normalization, exponent and threshold 
of the Kennicutt-Schmidt law at $z=0$,  
the values of $B/T$ do not change significantly 
when varying these parameters: slightly higher $B/T$ ratios are obtained for higher values of the normalization and the exponent, 
and lower values of the gas density threshold for SF. Only a very extreme deviation from the Kennicutt-Schmidt law 
at high redshifts would have a significant impact on the morphological mix of central galaxies at $z=0$. Another ingredient
in the satellite evolution is that of feedback-driven outflows, which are more effective for low-mass galaxies. 
If we remove this mechanism, the $B/T$ values increase in the low-mass end because SF is more efficient
in this case. At $10^9\msun$, a model without feedback predicts a 
fraction of bulgeless galaxies of $45\%$. At masses higher than $10^{11}\msun$, there are no changes in the morphological mix.

Finally, we note that in our fiducial model we have assumed that once the satellite enters the virial radius of the host, all its
hot gas corona, is immediately stripped. This is not
realistic for the most massive satellites that will be able to retain some of its gas for some time after accretion. If the cooling
time of this hot gas is lower than the merging time, then the gas will condense to the center of the satellite and contribute to the 
cold gas mass and also to SF. Although without a proper model, it is not possible to know the impact to our results, 
we speculate that satellites, specially the massive ones, will likely be more destructive resulting in a morphological
mix with more bulge-dominated galaxies at high masses.

\bibliography{./lit}

\label{lastpage}

\end{document}